\documentclass[dvips,11pt,a4paper]{article}
\usepackage[dvips,colorlinks,bookmarksopen,bookmarksnumbered,citecolor=red,urlcolor=blue]{hyperref}

\usepackage[T1]{fontenc}         
\usepackage[french]{babel}         
\usepackage[latin1]{inputenc}
\usepackage{graphicx}
\usepackage{amsmath}
\usepackage{bm}
\usepackage{tikz}
\usepackage{color}
\usepackage{helvet}
\usepackage{courier}
\usepackage{makeidx}
\usepackage{footmisc}
\usepackage{type1cm}         
\usepackage{textcomp}
\usepackage{subfig}
\usepackage{enumitem,xcolor}    

\DeclareCaptionLabelSeparator{colon}{. }
\usepackage[labelfont={bf},labelsep={colon}]{caption}

\DeclareMathAlphabet\mathbfcal{OMS}{cmsy}{b}{n}

\begin{document}
\thispagestyle{empty}

\begin{center}
{ \Large  \bf Extension of Discrete Mechanics towards Electromagnetism  } \\
\vspace{3.mm}
{\bf Jean-Paul Caltagirone } \\
\vspace{3.mm}
{ \small Université de Bordeaux  \\
   Institut de Mécanique et d'Ingéniérie \\
   Département TREFLE, UMR CNRS n° 5295\\
  16 Avenue Pey-Berland, 33607 Pessac Cedex  \\
\textcolor{blue}{\texttt{ calta@ipb.fr }  } }
\end{center}


The attempt to unify the laws of physics is approached from a discrete vision of space and time, abandoning the continuous medium paradigm that presided over the derivation of certain equations of physics - Navier-Stokes., Navier-Lamé, Maxwell, etc. Acceleration considered as an absolute quantity is expressed as a Hodge-Helmholtz decomposition, the sum of a solenoidal component and an irrotational component. Discrete mechanics, which has already unified mechanics in a relativistic formulation, is extended to electromagnetism with the same equation of motion expressed in terms of scalar and vector potentials. All the variables and parameters of this equation are described with only two fundamental units, those of length and time.
The discrete equation makes it possible to account for persistent phenomena in the absence of any excitation, permanent magnetization, pressure, shear stress and purely unsteady effects such as magnetic induction, longitudinal and transversal  wave propagation, differential rotation.


\vspace{3.mm}

\normalsize


\section{Introduction}
\label{intro}

The dynamic equations of J.C. Maxwell \cite{Max65}, also called Maxwell-Lorentz equations, are fundamental laws of physics. They constitute the basic postulates of electromagnetism, with the expression of the Lorentz's electromagnetic force. Today, Maxwell's four vector equations can be written in integral form, in just two tensor equations, or even in a single multivectorial equation in geometric algebra.
The application of the discrete exterior calculus \cite{Des05} to the equations of mechanics and those of electromagnetism made it possible to federate the discrete approach without unifying these laws. 



An example is the work of E. Tonti \cite{Ton13}, which reverses the approach by first using the discrete constitutive laws to truly define a unified approach.
Ideas evolve over time and the concepts of Galileo or Newton have led to modern formulations that have gone far beyond the scope of the concepts that led to their development. For example, newtonian mechanics is too often associated with an infinite wave speed, which is no longer the case since Euler derived the equations of the mechanics of compressible fluids. The dynamic vision of equations, adopted by Maxwell for electromagnetism, is essential, and introduces two important quantities for all physics, acceleration and velocity; the first can be considered absolute and the second is of course relative. 
The limitation of material velocity to the celerity of light is a postulate of relativistic mechanics which is not a prerequisite in discrete mechanics. This is the point of view adopted where, if there were any limit on velocity, it would be fixed by the decrease in acceleration when it approaches celerity, velocity and celerity being two disjointed notions which cannot be compared.

The unification of the physics continuum associating the constitutive laws of the mechanics of deformable media and electromagnetism is a major objective addressed in particular by A.C. Eringen et al. on electrodynamics of continua \cite{Eri90}. The goal was not the juxtaposition of the two theories but the search for a real composition. 
The authors establish a set of balance laws of electrodynamics of deformable bodies including Maxwell equations and momentum conservation laws and scan a wide spectrum of materials to which these equations apply.
The laws of the continuum physics thus obtained remain however in the classical framework of the continuous medium.

It is in this context that an attempt to unify the laws of mechanics was carried out in 2015 \cite{Cal15}; it led to the formulation of an equation of motion based on the weak equivalence principle (WEP) and the principle of relativity introduced by Galileo. The abandonment of the continuous medium paradigm made it possible to develop a discrete approach based on a quantity considered absolute, acceleration. The equation of the discrete movement resulting from this formulation covered all the properties of the Navier-Stokes equations for fluids, Navier-Lamé for solids, and a first result of General Relativity.

The physics underlying the Navier-Stokes, Navier-Lamé, General Relativity and Maxwell equations is the same; only the physical parameters are different. Since the time constants associated with these domains can be very different, it is possible that some effects are negligible and are not taken into account in the modelling of the phenomena; for example, classical mechanics does not need to use relativistic mechanics for applications where the velocity is much lower than the celerity of light in a vacuum. However, the problem here is to find a mathematical formulation to represent the largest number of physical phenomena for various spatial and temporal scales. Quantum mechanics is outside the scope of this paper, even though its contributions to the understanding of our usual physics is considerable, for example the duality between wave and corpuscle.
The aim here is to extend the concepts used to unify mechanics and electromagnetism without modifying the previously derived formulation, is not to find any analogy between mechanics and electromagnetism \cite{Mar98}, \cite{Rap14} nor even to introduce a new numerical methodology, but to formulate an equation representative of all these phenomena. The purpose is to derive a generic equation based on unique variables for all phenomena.

The velocity $\mathbf V$ considered here is that of fluid in motion, that of a solid loaded in stress or displacement, or the velocity of displacement of the electric charges in a conductor; for a copper conductor the velocity is of the order of $\mathbf V = 10^{-3} \: m / s$ but in a vacuum tube it is of the order of $\mathbf V = 10^{7} \: m / s$, a value close to the celerity of light.
Velocity itself is a secondary variable, it will appear as a lagrangian in equations updated by acceleration.
To clarify the discrete approach one can present it in a trivial way: a flux of particle or matter moves along a curvilinear trajectory with velocity $\mathbf V$ in a medium which can be a fluid, a solid or a vacuum and where celerity $ c $ of this medium is simply known. The particle, whether or not it has a mass, or the material, can be accelerated on its trajectory, thus modifying the velocity which becomes $\mathbf V = \mathbf V^o + dt \: \bm \gamma$ after a lapse observation  time $dt$. The acceleration is due to a certain number of physical phenomena, some of whose origins have not yet been understood in detail: gravitation, inertia, viscous friction, cohesion, etc. Only knowledge of acceleration $\bm \gamma$ and that of the previous state, defined by velocity $\mathbf V^o$, serve to define the current state. These quantities involve only two scales, that of time and that of space. The goal is to derive an equation of motion with variables that are characterized by only these fundamental units.

Discrete mechanics \cite{Cal15} has already been validated for the fields of fluid mechanics, solids or certain cases of General Relativity. It will be confronted here with the field of electromagnetism without any modification to the formulation. The aim is not to find analogies with Maxwell's equations but to extend discrete mechanics to discrete electromagnetism, with the absolute constraint of covering the laws of current electromagnetism in their field of application.

Other fields of physics are accessible by the proposed formalism. This is the case of the heat transfer where the heat flux vector and the temperature are respectively replaced by the velocity and the scalar potential; At the very low time constants we find the hyperbolic law of Cattaneo-Vernotte \cite{Ver61}. The mass transfer in the multicomponent mixtures or the flows in porous media correspond to applications treated in discrete mechanics \cite{Cal15}.

The undulatory character assigned to certain areas of physics makes it possible to address complex phenomena such as physical optics, which represent wave interference and diffraction. Some composite materials or synthetic three-dimensional structures have similar properties of polarization and chirality. The metamaterials created by T. Frenzel and his team \cite{Fren17}, \cite{Cou17} have the property of generating a twist when subjected to compression. These unconventional behaviours, which can thus be created on demand, open up very interesting applications perspectives but also a path of research focused on mathematical modelling on equivalent properties. The DM is entirely based on the propagation of longitudinal and transverse waves which allows a potential unification with the electromagnetism. Cases of complex constitutive laws and metamaterials have not been addressed.
Appropriate constitutive laws are already available in the context of the continuum mechanics \cite{Mil07}.

\textcolor{blue}{\section{Discrete formulation of conservation laws} }

\textcolor{blue}{\subsection{Notion of discreet medium} }

The establishment of a unified formulation of physics requires the use of quantities that federate all the variables associated with the multitude of laws of physics. Some of these laws are of greater importance, such as conservation laws, and others are simple laws of behaviour or constitutive laws. The best-known laws of physics are those of Navier-Stokes for fluids, Navier-Lamé for solids, and Maxwell for electromagnetism. Some of them are wrongly attributed to a Galilean vision, such as Newtonian mechanics; apart from the very specific case of so-called incompressible approximation, the celerity of fluid or solid media is not considered infinite. These equations thus enter the relativistic framework and are supposed to be in agreement with General Relativity as well as Maxwell's equations written in space-time formalism.

The search for a generic equation that would unify mechanics for all media and the propagation of all kinds of waves poses the problem of the existence of this physical law. The equations developed over the past three centuries were intended to represent only part of the observed reality. Maxwell brought together in a single dynamic theory the laws of Faraday, Joule, Ampère, Gauss, etc., using the notion of electromagnetic field. Progress in quantum mechanics and the advent of the theory of relativity at the beginning of the last century enabled a move towards a coherent description of certain branches of physics. The vast majority of work has revealed the existing laws and completed them, such as relativity for Newton's law; however, the path is still long and inconsistencies remain, for example the difference in formulation between fluid mechanics and solid mechanics, despite the notion of continuous media mechanics which was supposed to unify mechanics. 

The vision proposed here is based on the existence of a single law for all the areas of physics mentioned above. This law is built on the two principles derived from Galileo's observations, the principle of relativity and that of equivalence of the inertial and gravitational masses. These two principles suggest the existence of an invariant absolute variable independent of any spatial reference. This quantity is the acceleration $\bm \gamma$ that a medium or a particle would have under the influence of the acceleration imposed on it. Given the principle of equivalence, the generic law is written:
\begin{eqnarray}
\displaystyle{ \bm \gamma = \mathbf g  } 
\label{loiphys}
\end{eqnarray}

This law conforms to Newton's second law $m \: \bm \gamma = \mathbf F$ but here the vector $\mathbf g$ is the set of forces per unit mass applied to the medium or the particle. The law (\ref{loiphys}) expresses the conservation of acceleration, it is the only physical quantity which satisfies the mathematical vectorial addition $\bm \gamma = \bm \gamma_1 + \bm \gamma_2 $.
This law (\ref{loiphys}) is the cornerstone of the discrete mechanics developed in recent years \cite{Cal15}.
The basic assumptions of discrete mechanics are simply recalled here:
\begin{itemize}[label=\textcolor{blue}{\small \textbullet}]
\item the acceleration of a particle or a medium is an absolute quantity;
\item velocity is not limited, velocity and celerity are two disjointed notions; the first is a property of the medium (matter, vacuum) and the second is a relative quantity which can be accumulated;
\item the equivalence of gravitational and inertial masses as well as relativity are two intangible principles;
\item there is a scalar potential $\phi$ and a vector potential $\bm \psi$ of the same quantity $\bm \gamma$, acceleration;
\item the Hodge-Helmholtz decomposition applies to the acceleration that breaks down into one irrotational component and another solenoidal component;
\item source terms resulting from physical effects such as inertia, gravity, capillarity, etc. are decomposable according to this same principle.
\end{itemize}

Any vector can be decomposed into an irrotational part and a null divergence part:
\begin{eqnarray}
\displaystyle{ \bm \gamma = - \nabla \phi + \nabla \times \bm \psi  } 
\label{dechh}
\end{eqnarray}

The two fields $\nabla \phi$ and $\nabla \times \bm \psi$ are orthogonal. We will adopt the principle that any vector can be written in this form. This decomposition is sometimes presented with a third harmonic term, at once null divergent and null rotational; in fact, this term is closely associated with the uniform overall movements that must disappear from the formulation under the principle of equivalence. Velocity is a variable whose absolute value does not matter, in discrete mechanics it is considered as a simple lagrangian upgraded by the acceleration $\mathbf V = \mathbf V^o + dt \: \bm \gamma $ where $\mathbf V^o $ is velocity at moment $t^o$ and $dt$ the time between two observations of the phenomenon.
\begin{figure}[!ht]
\begin{center}
\includegraphics[width=5.cm,height=3.cm]{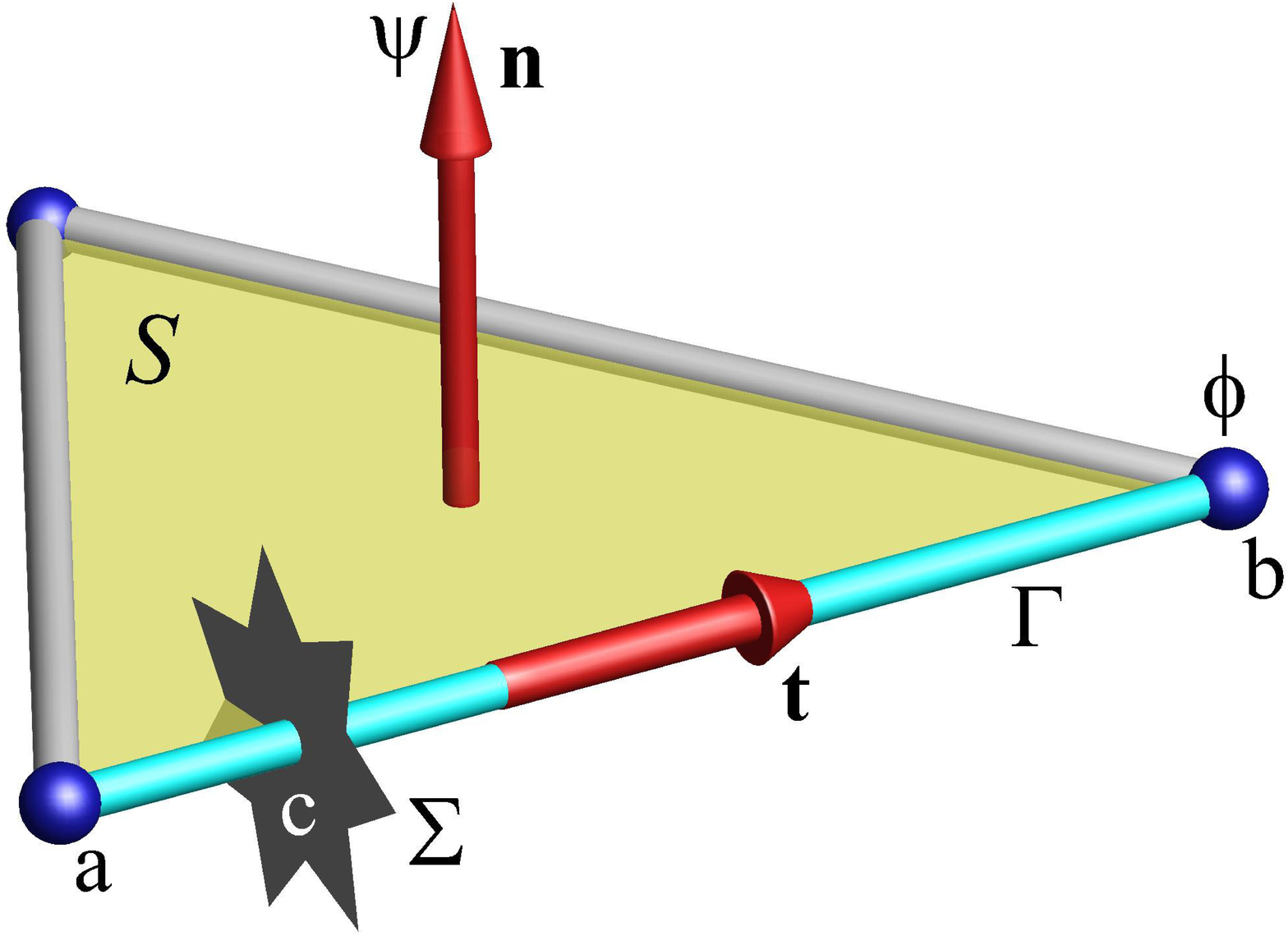}
\hspace{10.mm}
\includegraphics[width=4.6cm,height=5.cm]{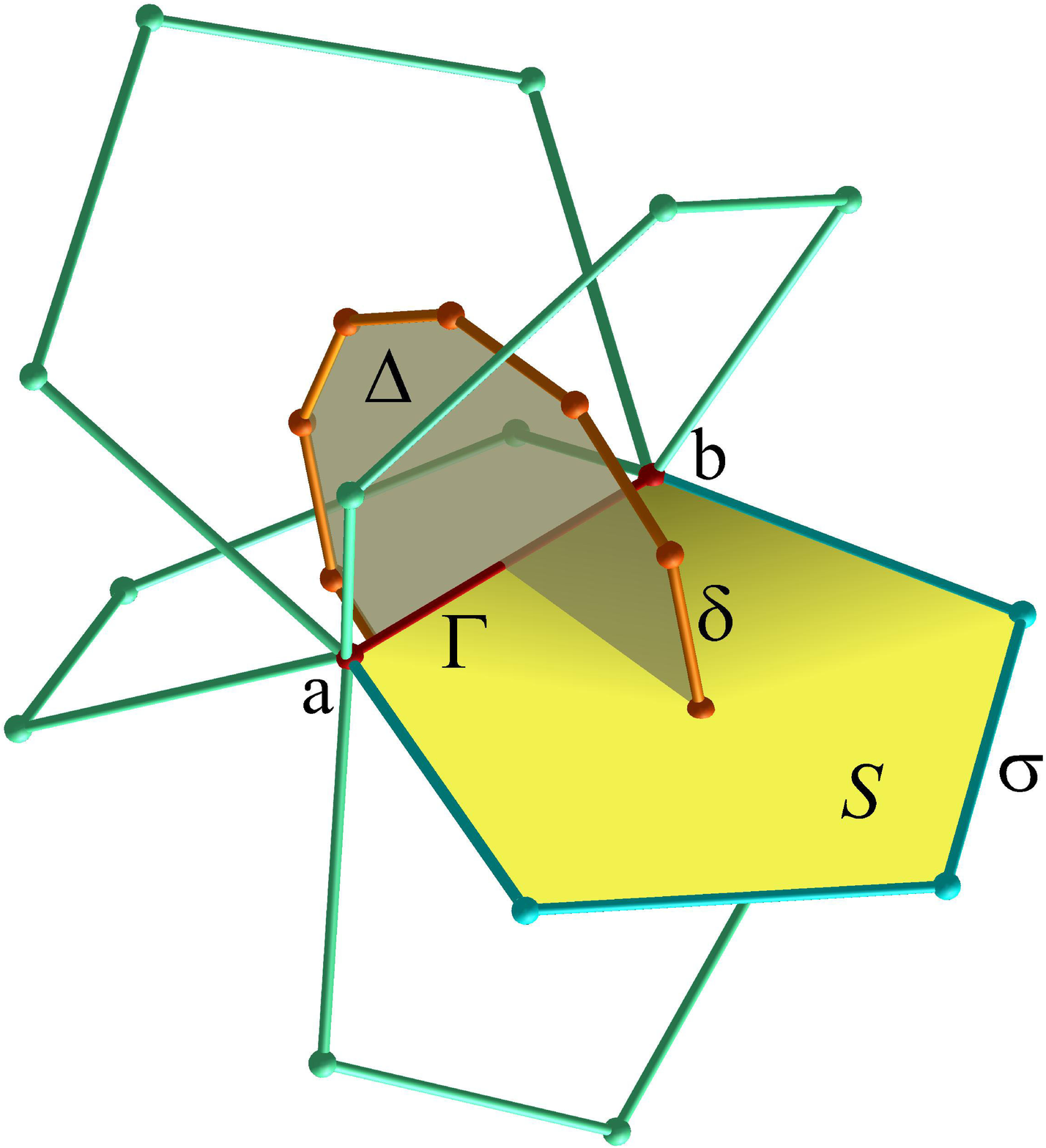}  \\
\vspace{-1.mm}
\hspace{-1.mm} (a) \hspace{60.mm} (b)
\caption[Elementary geometrical structure of discrete media mechanics]{\it (a) Elementary geometrical structure of discrete media mechanics: three straight $\Gamma$ edges delimited by dots define a planar face $\mathcal S$. The unit normal vectors $\mathbf n$ to the face and the vector carried by $\Gamma$ are orthogonal, $\mathbf t \cdot \mathbf n = 0$. The edge $\Gamma $ can be intercepted by a discontinuity $\Sigma$ located in $c$, between the ends $a$ and $b$ of $\Gamma$. $\phi$ and $\mathbf \Psi$ are the scalar and vector potentials respectively. (b) The virtual machine of motion in Discrete Mechanics: the acceleration of the medium along the edge $\Gamma$ is due to the difference of the scalar potential $\phi$ between the ends of the edge $[a, b]$ of unit vector $\mathbf t$, to the circulation action of the vector $\mathbf V$ on the contour of the different primal facets $ \mathcal S $ inducing an acceleration on $\Gamma$ and the projection $\mathbf g \cdot \mathbf t $ imposed other accelerations as gravity.}
\label{discreet}
\end{center}
\end{figure}

Acceleration is thus written as the sum of the two terms $\bm \gamma = \bm \gamma_{\phi} + \bm \gamma_{\psi}$ which can be modelled according to velocity and medium properties.
Velocity $\mathbf V$ also has two components that are upgraded by the components of acceleration $\mathbf V_ {\phi} = \bm \gamma_{\phi} \: dt$ and $\mathbf V_{\psi} = \bm \gamma_{\psi} \: dt$ which represent the currents flowing over the segment $\Gamma$; as the two fields are orthogonal, these currents do not have any direct interaction.
In fact it is not possible, in general, to extract the components of velocity $\mathbf V$ by directly applying a local discrete Hodge-Helmholtz decomposition, the result depends closely on the boundary conditions. Fortunately the $\mathbf V$ field is relative and is only a secondary variable; it is the decomposition of acceleration $\bm \gamma$, the absolute quantity, which is sought in the form of potentials; and this decomposition is possible.

The notion of continuous medium is also abandoned, as is that of global reference; there is a local discrete geometrical structure represented in the figure (\ref{discreet}) composed of a primal and a dual elementary structures. 
The term discrete geometric structure or geometric topology used defines a set of links between connected elementary objects, here points, segments, surfaces. These notions are identical to those that can be found in mesh structures, for example resulting from spatial discretization in finite elements.
The segment $\Gamma$ of unit vector oriented $\mathbf t$ and of ends $a$ and $b$ defines the basic element of the primal topology which, with two other segments, forms the planar surface $\mathcal S$ whose unit-oriented normal is $\mathbf n$ such that $\mathbf t \cdot \mathbf n = 0$ (figure \ref{discreet}(a)). The scalar potential $\phi$ is only defined on the ends of the primal topology. A possible contact discontinuity or shock wave $\Sigma$ intersects the segment $\Gamma$ into $c$. The normal to the $\mathcal S $ plane is associated with a pseudo-vector $\bm \psi$ such that the rotation of vector $\mathbf V$ is itself associated with the segment $\Gamma$. Figure (\ref{discreet}(b)) represents the primal surface $\mathcal S$ as a planar polygon; the $\delta$ outline and the $\Delta$ surface form the dual topology.

The presentation of the differential operators may be greatly different depending on the degree of formalization of the differential geometry \cite{Mar02}. The succinct and non-exhaustive presentation given here is based on a simple physical approach which allows us to define the operators associated with the switch from one topology to another, on the basis of scalar or vectorial information. It should be remembered that although the classic notion of a continuum has been set aside, the material is a continuum where the directions of the edges and of the normal to the surfaces are also preserved at all scales of observation.

 The gradient operator applied to a scalar $\phi$, $\nabla \phi$, represents the difference of that scalar over a distance $d$ between $a$ and $b$ in a given direction. Unlike with the concept of continuum mechanics, the gradient vector defined here has only one component, assigned as a scalar to the edge $\Gamma$. The gradient of a scalar in the space has no meaning - it is an illegal operation in discrete mechanics. The gradient is calculated solely on a bipoint $[a b]$ linked by an edge.
 The primal rotational of a vector $\mathbf{W}$, $\nabla \times \mathbf{W}$, is associated with the circulation of its components $\mathbf{V}$ over all the edges $\Gamma$ constituting the primal boundary. It is represented by a vector $\mathbf{n}$ orthogonal to the primal surface. This surface is considered to be flat. This apparent restriction disappears as the surface area $ds$ tends toward zero; however, it will remain a condition for the application of the theorems of differential geometry in particular context discussed here.  
 The divergence represents the flux of a vector $\mathbf{W}$, $\nabla \cdot \mathbf{W}$, across all the facets of
the dual surface. The scalar obtained as assigned to the single point inside the dual volume. The flux is calculated on the basis of the components $\mathbf{V}$ on the edges $\Gamma$ of the vector itself. If the vector $\mathbf{W}$ is a rotational, calculated as the circulation of another vector on each primal boundary, then the divergence will be strictly null. 
The dual rotational of a vector $\mathbf{W}$, $\nabla \times \mathbf{W}$, physically represents the flux of the vector $\mathbf{W}$ across that portion of the dual surface associated with the edge $\Gamma$. This flux is calculated using the circulation of the vector, or rather, of its components, on the boundary delimiting the dual portion which, in general, is not flat. The result of this operation is assigned to the edge $\Gamma$ as a vector or a directed scalar, if necessary.

The circulation of vector $\mathbf V$ along the contour of the primal surface makes it possible to calculate the primal rotation of unit vector $\mathbf n$, and the circulation along the $\delta$ contour of the surface $\Delta$ is the dual rotational that re-projects the result on the $\Gamma$ segment. Note that the number of planes of normals $\mathbf n$ associated with the segment $\Gamma$ is arbitrary, five in the figure (\ref{discreet}(b)). In the context of a continuous medium, the requirement to use a reference system leads us to consider the three components of the vectors, the nine components of a second-order tensor, and so on. The same reasons require the introduction of the four components of space-time, the fourth order Riemann tensor in relativity theory. The discrete description serves to satisfy, from the outset, the notion of material indifference and to represent the notion of polarization.

The primal and dual topologies thus defined serve to satisfy two essential properties in discrete $\nabla_h \times \nabla_h \phi = 0$ and $\nabla_h \cdot (\nabla_h \times \bm \psi) = 0$, whatever the topologies based on planar surfaces, polygons or polyhedra and whatever the regular functions $\phi$ and $\bm \psi$. These conditions are absolutely necessary for a complete Hodge-Helmholtz decomposition applied here to acceleration. Each vector can be decomposed into a solenoidal part and an irrotational part, but the scalar and vectorial potentials are not of the same importance according to the nature of the vector. In electromagnetism $\phi$ and $\mathbf A$ do not come from the same vector and do not express themselves with the same units. In mechanics the scalar potential and the velocity vector potential do not have any particular physical importance but can be used to project velocity over a field with zero divergence. Only acceleration $\bm \gamma$ and its potentials $\phi$ and $\bm \psi$ have special physical properties.

It should be noted that $\phi^{o}$ and $\bm{\psi}^{o}$ are the stresses at time $t$, where all the forces applied before that instant are ``remembered''. The formalism presented here enables us to take account of the entire history of the medium, i.e. its evolution over time from an initial neutral state. For a given instantaneous state of strain, there may be multiple paths by which that state can be reached, and $(\phi^{o},\bm{\psi}^{o})$ will, alone, contain the whole of the medium\textquoteright s history. It is not helpful to know the local and instantaneous stresses, in that these two potentials will have accumulated stresses over time; these quantities are also called ``accumulators'' or ``storage potentials''. These potential can therefore be used to take account of the behaviour of media with continuous memory.

\textcolor{blue}{\subsection{Discrete motion equation} }

The derivation of the equation of discrete motion is essentially based on the physical modelling of phenomena observed in both mechanics and electromagnetism. The elementary laws representing these phenomena exist, they are most often written as linear relations between flows and forces. Some of them may have different representations in CM and DM, for example the viscous effects interpreted as shear and written in the form of a CM derivative can result in a primal rotation in DM \ cite {Cal15}. Whatever the physical phenomenon observed, it can be modelled in DM only with two types of effects, a longitudinal directional effect linked to the longitudinal celerity of the medium $c_l$ and another transversal associated with the transverse celerity $ c_t $ possessing the property of polarization. In mechanics, for example, compression is described by the product $c_l \: \nabla \cdot \mathbf V $ and the shear stress by $c_t \: \nabla \times \mathbf V $. In electromagnetism the velocity is replaced by a function of the electric current. The derivation of the discrete equation is directly carried out on the segment oriented $\Gamma $ considering that the intrinsic acceleration of the medium $\bm \gamma$ is equal to the sum of the external accelerations $\mathbf g$.

The equation of motion is established for all media, fluids, solids or vacuum. Properties can include:
\begin{itemize}[label=\textcolor{blue}{\small \textbullet}]
\item fluids can be compressible or incompressible, Newtonian or complex rheologies, nonlinear, viscoelastic, viscoplastic, with thresholds, etc;
\item solids can correspond to various constitutive laws, from the elastic solid to the laws of complex behaviours; the unsteady temporal processing of the model makes it possible to treat cases of large deformations and large displacements while conserving the mass;
\item electromagnetic media can have any properties corresponding to those of fluids, solids or vacuums; persistent states such as electric charge accumulation, permanent magnetization or hysteresis effects are implicitly incorporated into the formulation.
\end{itemize}

The derivation of the previously realized equation of motion \cite{Cal15} leads to the system of generic equations of discrete mechanics:
\begin{eqnarray}
\left\{
\begin{array}{llllll}
\displaystyle{ \bm \gamma = - \nabla \left( \phi^o - dt \: c_l^2 \: \nabla \cdot \mathbf V \right) + \nabla \times \left( \bm \psi^o - dt \: c_t^2 \: \nabla \times \mathbf V \right) - \kappa \: \mathbf V + \mathbf g } \\  \\
\displaystyle{ \phi  = \alpha_l \: \phi^o - dt \: c_l^2 \: \nabla \cdot \mathbf V } \\ \\
\displaystyle{ \bm \psi  = \alpha_t \: \bm \psi^o - dt \: c_t^2 \: \nabla \times \mathbf V } \\ \\
\displaystyle{ \mathbf V = \mathbf V^o + \bm \: \gamma \: dt } 
\end{array}
\right.
\label{mecadis}
\end{eqnarray}

Before giving an expression of the stress state, it is advisable to define mechanical equilibrium. Mechanical equilibrium is obtained when the law of dynamics is satisfied, $\bm{\gamma}-\mathbf{g}=0$ if $\mathbf{g}$ represents all the forces  per inut mass applied to the system studied at time $t$. This choice of the concept of equilibrium precludes all motions where acceleration is null, the uniform rectilinear motion.

Any disturbance to this state of equilibrium due to modifications made to one of the source terms, the boundary conditions, etc., will lead to a change in the variable used - i.e. velocity or displacement - which leads the system to a different state of equilibrium at time $t+dt$, for which we shall also have $\rho\:\bm{\gamma}-\mathbf{f}=0$. What changes between the two states of equilibrium is the residual stress state manifested by two potentials $\phi^o$ and $\bm \psi^o$. Hence, for a state of mechanical equilibrium, the sum of the relative contributions to these two values is null and the acceleration is also null.

The quantity $\phi^o$ is the equilibrium scalar potential and $\bm \psi^o$ is the equilibrium vectorial potential. They express the persistence of a physical phenomenon such as electric potential and permanent magnetization in electromagnetism or pressure and shear stress in a solid medium in mechanics. In the absence of any movement of the particle or the medium and in the presence of a source term $\mathbf g$ the equation becomes $- \nabla \phi^o + \nabla \bm \psi^o + \mathbf g = 0$. The factors $\alpha_l$ and $\alpha_t$ are physical quantities to express the persistence of long-term effects. For example, the relaxation time of shear stresses in a fluid medium is of order of magnitude of $\tau_f \approx 10^{-12} \: s$ and the $\alpha_t$ factor can legitimately be set to zero for common applications. The quantity $\nabla \phi$ is a polar vector just like $\nabla \times \bm \psi$ whereas $\bm \psi$ is an axial vector or pseudo-vector.

The parameter $- \kappa \: \mathbf V$ represents the effects of viscous friction on a small scale; in fluid mechanics this term is the volume drag of Darcy $- (\mu_f / K) \: \mathbf V $ where $\mu_f$ is the viscosity of the fluid and $K$ is the permeability of the porous medium. In electromagnetism this term enables to modelling the reduction of acceleration of electric charges.

Longitudinal $c_l$ and transverse $c_t$ celerities are intrinsic properties of the medium, matter or vacuum; these quantities depend on multiple variables but they will be assumed simply to be known in space and time. The definition of these celerities also depends on the physical phenomenon studied - electromagnetism, fluid mechanics, etc. Velocity $\mathbf V$ is a quantity that is totally independent of celerity, the two notions are strictly disjointed and cannot be compared as in Special Relativity. The duality between wave and particle introduced in quantum mechanics makes it possible to assign a celerity to the wave and a velocity to the particle. If the velocity happens to be limited by a value equal to the celerity, it means that the acceleration tends towards zero. Velocity is a secondary variable, a lagrangian, which is updated from the acceleration and time $ dt $ between two observations of the physical system.

The system (\ref{mecadis}) is composed of a vector equation that calculates the variable $\mathbf V$ and three updates of potentials and velocity. Acceleration $\bm \gamma$ can be replaced by the material derivative definition $\bm \gamma = d \mathbf V / dt$ to give an implicit law. The material derivative itself can be replaced by expressing the terms of inertia $\bm \gamma_i = \partial \mathbf V / \partial t - \nabla \times (| \mathbf V |^2/2 \: \mathbf n) + \nabla (| \mathbf V |^2/2) $. The form of the inertial terms discussed in \cite{Cal15} is applicable whatever the medium considered: fluid, solid, vacuum.

Similarly, all possible source terms $\mathbf g$ are considered accelerations and thus decomposed according to the Hodge-Helmholtz form. This is the case of gravity where the scalar potential is $\phi_g = \mathcal G \: M / r$ thus giving the two contributions of gravitational acceleration $\bm \gamma_g = - \nabla \phi_g + \nabla \times \phi_g \: \mathbf n$; another example is that of capillary acceleration which is written $\bm \gamma_c = - \nabla \phi_c + \nabla \times \phi_c \: \mathbf n $ with $\phi_c = \sigma \: \kappa$ where $\sigma$ is the surface tension per unit mass and $\kappa$ the longitudinal or transverse curvature.

The equation of the system (\ref{mecadis}) is written with quantities that are expressed only with the fundamental units, length and time, whereas the dedicated equations, Navier-Stokes, Navier-Lamé and Maxwell involve the set of fundamental units, length $L$, time $T$, mass $M$, intensity $I$ and sometimes temperature $\Theta$.

 For the phenomena expected to be described by equation (\ref{mecadis}), i.e. mechanics of fluids and solids, relativistic mechanics and electromagnetism, table (\ref{corres}) presents the correspondence between the variables and the properties conventionally used for each of the domains described and those to be fixed in the equation, $dt \: c_l^2$ and $dt \: c_t^2 $. The quantities present in this table are respectively $\chi_T$ the coefficient of isothermal compressibility, $\mu_f$ the viscosity of the fluid, $\nu_f$ its kinematic viscosity, $\lambda$ and $\mu_s$ the Lamé coefficients, $\rho_m$ electrical density, $\varepsilon_m$ permittivity, $\mu_m$ magnetic permeability and $\sigma_m$ electrical conductivity. The variable of the discrete equation is the velocity $\mathbf V$ of the particle or the medium, $\mathbf U$ is the displacement of the solid, $e$ is the electric potential, $\bm j$ is the density of electrical current, and charge density is noted $\rho_m$.

It is easy to extract from this table (\ref{corres}) all the electromagnetic quantities from the potential ones $\phi$, $\bm \psi$ and the velocity $\mathbf V$; thus the rotational of the vector potential is equal to $\nabla \times \bm \psi = \nabla \times \left((\rho_m / (\rho \: \sigma \: \mu_m)) \: \mathbf B \right) $ and we find the Maxwell-Thomson law $\nabla \cdot \mathbf B = 0$, i.e. the fact that the magnetic field has no charge. As in mechanics, the physical quantities are integrated within the operators gradient and rotational dual and they do not have to be derived in space, they are constant on the whole primal surface $\mathbf S$. It can be seen that the correspondence between mechanical and electromagnetic variables has nothing in common with the analogies presented in the literature. In discrete mechanics the equation is the same and the potentials of a true Hodge-Helmholtz decomposition are those of a single quantity, acceleration.
\begin{table}[!ht]
\begin{center}
\begin{tabular}{|c|c|c|c|c|c|c|}   \hline
          &   $\mathbf V$  &$\phi$  &   $\bm \psi$      & $dt \: c_l^2 $   &  $dt \: c_t^2 $  \\ \hline  \hline
fluids & $\mathbf V$  & $ p / \rho $ & $\bm \omega / \rho$ & $dt / ( \rho \: \chi_T ) $   & $ \nu_f = \mu_f / \rho $  \\ \hline
solids  & $\mathbf U /dt $  & $ p / \rho $ & $\bm \omega / \rho$ & $dt \: (\lambda + 2 \: \mu_s) / \rho $ & $\nu_s = \mu_s / \rho $ \\ \hline
électro.    & $\bm j / \rho_m$  & $ (\rho_m  /  \rho) \: e$  & $(\rho_m / (\rho \: \sigma \: \mu_m)) \: \mathbf B$ & $dt / ( \varepsilon_m \: \mu_m ) $   & $\nu_m = 1/ (\mu_m \: \sigma_m) $  \\ \hline
\end{tabular}
\caption[Correspondence]{ \it Correspondence of quantities, variables and properties, used in discrete mechanics and the usual quantities in mechanics and electromagnetism where $\mathbf V$ is velocity, $\mathbf U$ displacement, $p$ pressure, $\bm \omega$ the constraint, $\bm j$ the current density, $\rho_m$ the electrical charge density and $\mathbf B$ the induction magnetic field. }
\label{corres}
\end{center}
\end{table}

The density $\rho$ is that of the fluid medium $\rho_f$ or the solid $\rho_s$ or the vacuum $\rho_v = 0$; even in the latter case the potentials continue to make sense. For example, for a perfect gas $\phi = p / \rho = r \: T$ continues to have a value as long as the notion of temperature continues to make sense. For the electromagnetic phenomena in the vacuum it is the current $\bm j$ which becomes null at the same time as the density. These cases correspond to very compressible media for which the divergence of the velocity is very important; deletions of terms {\it a priori} are to be applied with a great care, as the product of two terms, one of which tends to zero and the other to infinity, is of course undetermined. The system (\ref{mecadis}) is unsteady and applies in all the cases previously mentioned whatever the time-lapse $dt$ considered: for values compatible with the physics of phenomena, including the propagation of light, the system will report evolutions in time. For much larger values of $d t$ the evolution will not be physical but the convergence state will correspond to the stationary solution of the problem. Given the highly implicit character of the discrete equation, the formulation is very robust.

In electromagnetism the velocity in the vacuum $c_0$ is equal to the celerity of light and there is no transverse celerity. There are, however, polarizable transverse gravitational waves whose velocity is currently fixed at longitudinal velocity $c_0$; in the absence of different information we will use this result. It is recalled that the velocity of the particles (matter, electrons, photons) is not limited and that the celerities of the associated waves are exclusively measured quantities. The quantity $p^o$ is the equilibrium mechanical pressure and $\bm \omega^o$ the perfectly defined shear-rotation stress in fluid and solid.

The factors $\alpha_l$ and $\alpha_t$, which are dimensionless quantities between $0$ and $1$, are also intrinsic properties of the media. These factors depend largely on the time constants $\tau_f$, $\tau_s$ and $\tau_m$ corresponding to the relaxation times of the transverse phenomena.
For example, for water, if $\nu_f = \mu_f \: \rho \approx 10^{-6}$ is the kinematic viscosity and $c_l^2 = 1 / (\rho \: \chi_T) \approx 2.25 \: 10^6$, the characteristic time is then $\tau_f = \nu_f / c_l^2 \approx 10^{-12} s$. It should be noted that transverse celerity is not known for water. It is understood that water relaxes the shear stresses on characteristic times greater than $\tau_f$, and we can adopt $\alpha_t = 0$. For times $dt$ of this order of magnitude, the accumulation of shear stresses is no longer negligible. The same analysis can be made for the accumulation of constraints in a solid where we have $\nu_s \approx 2 \: 10^7$ and $c_l^2 \approx 5 \: 10^6$ that is $\tau_s \approx 0.25$ in copper; in this case copper accumulates the shear stresses $\alpha_t = 1$. For dielectric materials the values are very variable and it is necessary to perform a preliminary analysis.

The transformation of mechanical or electrical energy into heat is due to the viscous friction described by the rotational dual  but also by the term $- \kappa \: \mathbf V$ of the equation (\ref{mecadis}). This dissipation is evaluated by the function $\Phi_d = dt \: c_l^2 \: (\nabla \cdot  \mathbf V)^2 + dt \: c_t^2 \: (\nabla \times \mathbf V)^2 + \kappa | \mathbf V |^2$ in discrete mechanics. In electromagnetism this last contribution corresponds to the Joule's law which is written in this context $- \nabla \phi \cdot \mathbf t = \kappa \: | \mathbf V |^2$ by linking the potential difference to the dissipation.

Contrary to what one might think, one cannot eliminate terms with very large or very small coefficients. For example, if we want the flow to be incompressible, we must keep $\nabla \cdot \mathbf V$ in the equation of motion and in fact it is when the longitudinal velocity is very high that the divergence becomes very low, the term $dt \: c_l^2 \: \nabla \cdot \mathbf V$ is an order of magnitude of the other terms of the equation, {\it a priori} of order one. These factors make it possible to maintain persistent effects in the absence of any velocity, such as the magnetic field of a magnet in the very long term. Its demagnetization and the hysteresis induced by a current will naturally be taken into account by the equation of motion.

The system (\ref{mecadis}) and the properties (\ref{corres}) are sufficient to deal with any problem in one of the domains mentioned. The variables are the scalar potential $\phi$ defined at the ends of the segments $\Gamma$ and the vector potential $\bm \psi$, a pseudo-vector, associated with the normals $\mathbf n$ of each of the primal faces; the number of facets having the $\Gamma$ segment in common is variable and the vector, tensor or quadrivector formulation no longer makes sense in this discrete context. It is of course possible to return to the usual variables, for example electric potential $e$, electric field $\mathbf E$, magnetic field $\mathbf B$, excitation field $\mathbf H$, magnetization $\mathbf M$, charge density $\rho_m$,  current density $\bm j$, etc. for electromagnetism where $c^2 = 1 / \varepsilon \: \mu $. All these quantities which are not independent have been defined over time and have become usual notions, but it is no less legitimate to consider only $\phi$ and $\bm \psi$.

It is necessary to add a law of conservation on a particular potential, mass for  fluids $\rho_f$ and solids $\rho_s$ and $\rho_m$ density of charge in electromagnetism; in discrete, the law of conservation is found for all cases treated:
\begin{eqnarray}
\displaystyle{  \frac{d \rho }{dt}  + \rho \: \nabla \cdot \mathbf V = 0 }
\label{masse}
\end{eqnarray}

In discrete, for a process of temporal accumulation between two states at the instants $t^o$ and $t^o + dt$, the states are expressed in the same way:
\begin{eqnarray}
\displaystyle{ \rho  = \rho^o - \rho^o \: dt \: \nabla \cdot \mathbf V }
\label{massed}
\end{eqnarray}

The equilibrium potentials $\phi^o$ and $\bm \psi^o$ correspond to the potential accumulation of longitudinal and transversal actions over time; these persistent terms are written:
\begin{eqnarray}
\displaystyle{ \phi^o  = \int_0^{t^o} c_l^2 \: \nabla \cdot \mathbf V \: dt ; \:\:\:\:\:\:\:\:\:\:\: \bm \psi^o  = \int_0^{t^o} c_t^2 \: \nabla \times \mathbf V \: dt}
\label{potequ}
\end{eqnarray}

The solution to any problem is to find $\left(\phi, \bm \psi, \rho \right)$ as a function of space and time. These quantities corresponding to each equilibrium, defined as the exact satisfaction of equation (\ref{mecadis}), are persistent, and stopping the integration process in time will not modify the values.
Physical properties are also updated if they depend on variables and time.

\textcolor{blue}{\section{Application to mechanics} }

The equation system of Discrete Mechanics (DM) (\ref{mecadis}) is representative, without modification, of several domains of physics, fluid mechanics, solid mechanics, wave propagation or heat transfer, to cite a few.

Numerous examples have shown the validity of system (\ref{mecadis}) in fluid mechanics. For example, compressible, incompressible, non-isothermal or two-phase flows can be approximated with DM to recover the classically known results, in particular the classical analytical solutions of Poiseuille or Couette flows. Synthesis solutions of the Navier-Stokes equation, such as the Green-Taylor vortex, are also valid and satisfied by the DM equations. Reference cases, such as the lid-driven cavity, the backward facing step or the flow around a cylinder, make it possible to show that the DM model converges to order two in space and time for both velocity and pressure. The flows associated with heat and mass transfers including multi-components are reproduced in a similar way. More complex problems of shockwaves like the Sod tube, phase changes, boiling, condensation \cite{Ami14} are treated in a coherent way by integrating discontinuities within the equations of motion. Two-phase flows with capillary effects, surface tension or partial wetting, are particularly well suited to the discrete mechanics model \cite{Cal15c}.

In the present form, system (\ref{mecadis}) is relatively close of the Navier-Lam\'e equation associated to the study of stresses and displacements in solids. It differs, however, on several points: especially the discrete formulation is established in velocity, the displacement is only an accumulation of $\mathbf V \: dt$ as the velocity is itself the raising of $\bm \gamma \: dt $. Numerous examples of simple solicitations make it possible to find the solutions of the Navier-Lam\'e equation. More complex 2D and 3D problems on monolithic fluid-structure couplings have already made it possible to validate the proposed formulation \cite{Bor14}, \cite{Bor16}. The vision of a continuous memory medium makes it possible to treat the problems of large deformations and large displacements in a formulation where the pressure stress and the shear are obtained at once in a synchronous way without compatibility conditions. Given the original dissociation between compression effects and rotation, the material indifference introduced by Truesdell \cite{Tru74} is satisfied naturally. The complex constitutive laws can be treated without difficulty, only the physical parameters written in the equation of motion must be known.

The Special Relativity based on the concepts of Galileo, relativity and equivalence, allowed Maxwell, Poincar{\'e} and many others to introduce the Lorentz transformation into the equations, and in particular Einstein to make of equivalence notion a strong principle by adding the notion of velocity limit, that of light in vacuum. The ratio $v^2 / c^2 $ of the Lorentz transformation combines two quantities which are not of the same nature, the velocity is a relative quantity and the celerity a property of the medium. The velocity cannot be limited {\it a priori} by the celerity of the light. In fact, if the velocity should tend towards a limit, the reason for this asymptotic value is that the acceleration becomes zero. The system of equations (\ref{mecadis}) is perhaps relativistic, this is what is advanced in the present work. This statement is motivated by presenting how the discrete mechanics makes it possible to find the first major result of the theory of the relativity, the deflection of light by the Sun.

The two very simple cases that follow make it possible to illustrate the role of the set of scalar and vector potentials $ (\phi^o, \bm \psi^o)$ within the equation (\ref{mecadis}); the compression energy associated with $\nabla \cdot \mathbf V$ and rotation energy linked to $\nabla \times \mathbf V$ is stored and released at each instant by these potentials, which are energies per unit mass. The terms in gradient and in dual rotational are orthogonal and cannot exchange these energies directly, an imbalance thus leads to an acceleration, positive or negative. The accumulation of shear stresses in an elastic or viscoelastic solid leads to a state of rest if the boundary conditions are stationary. The mechanism itself of the propagation of longitudinal and transverse waves in a material is engraved in this formalism to describe these complex exchanges.

\textcolor{blue}{\subsection{A simple case} }

We consider one of the simplest cases of fluid-structure interaction to study the behaviour of two media, one viscous and the other elastic. This test case has a very simple analytical solution that highlights the behaviour of the two media modeled with the discrete description (\ref{mecadis}). The domain height $h = 1$ is separated by a $\Sigma$ interface located at $h / 2$. The velocity of the lower wall is kept at rest and the upper surface is initially set in motion with a velocity $V_0 = 1$.
\begin{figure}[!ht]
\begin{center}
\includegraphics[width=4.3cm,height=4.cm]{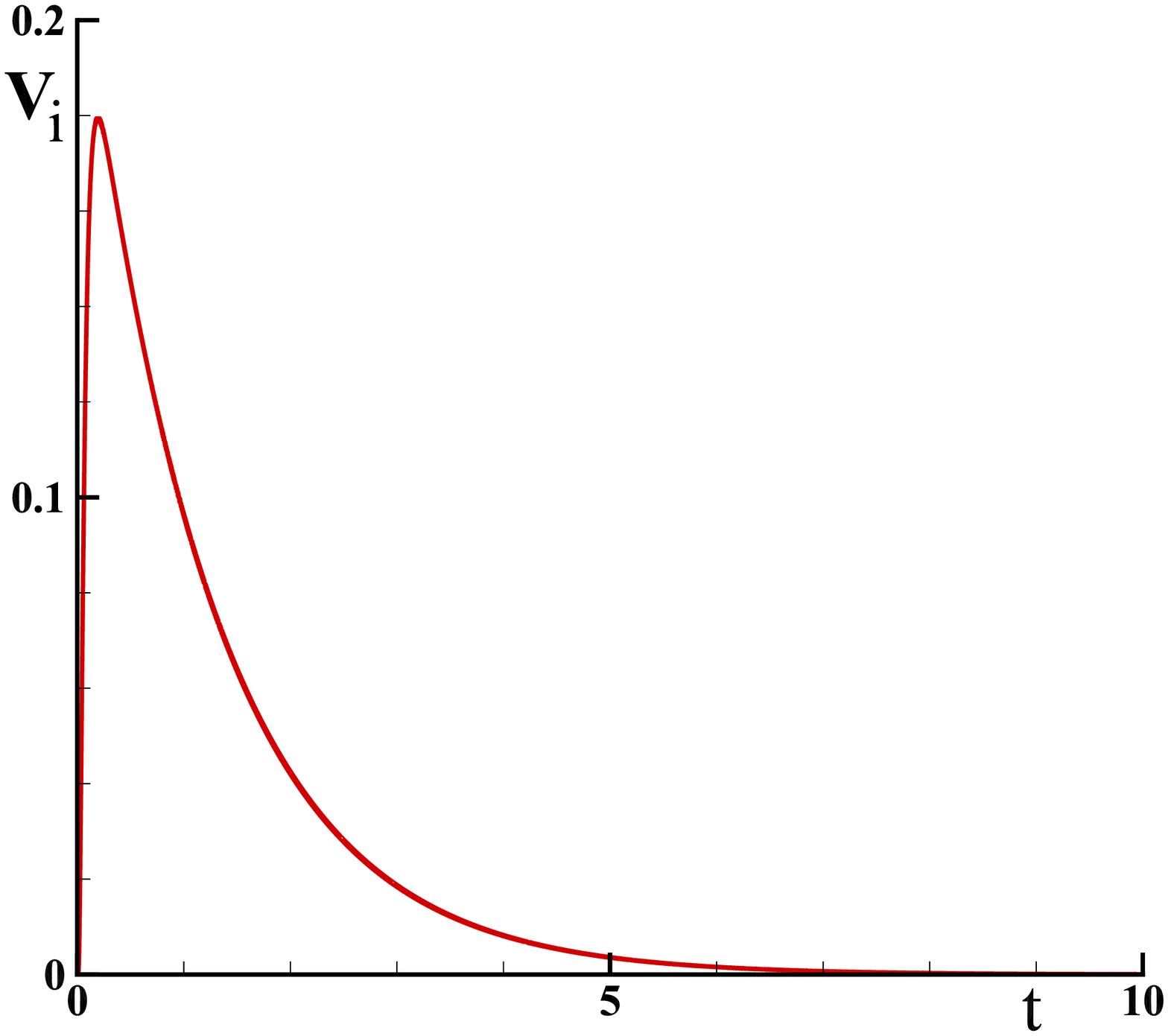}
\includegraphics[width=4.3cm,height=4.cm]{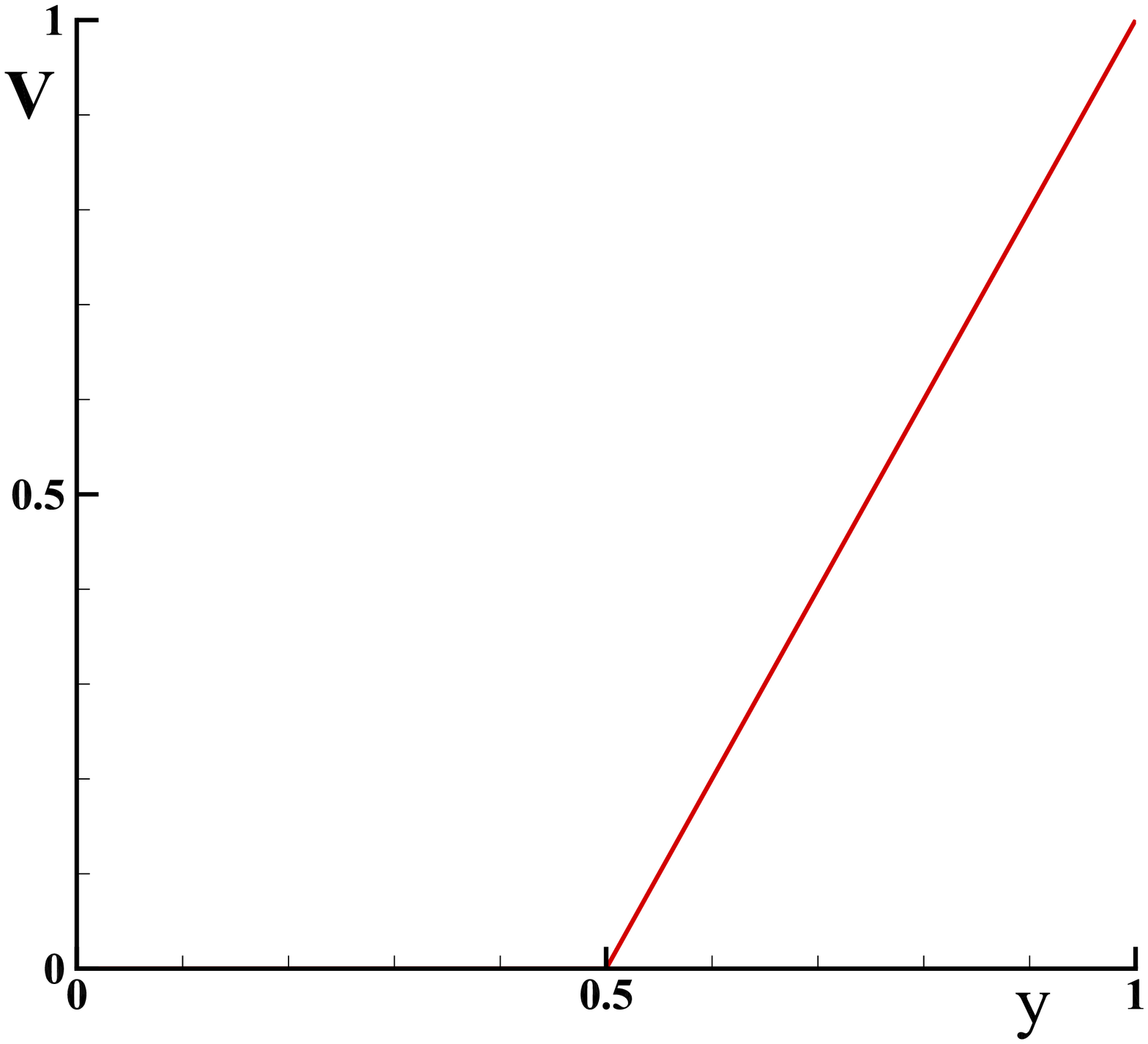}
\includegraphics[width=4.3cm,height=4.cm]{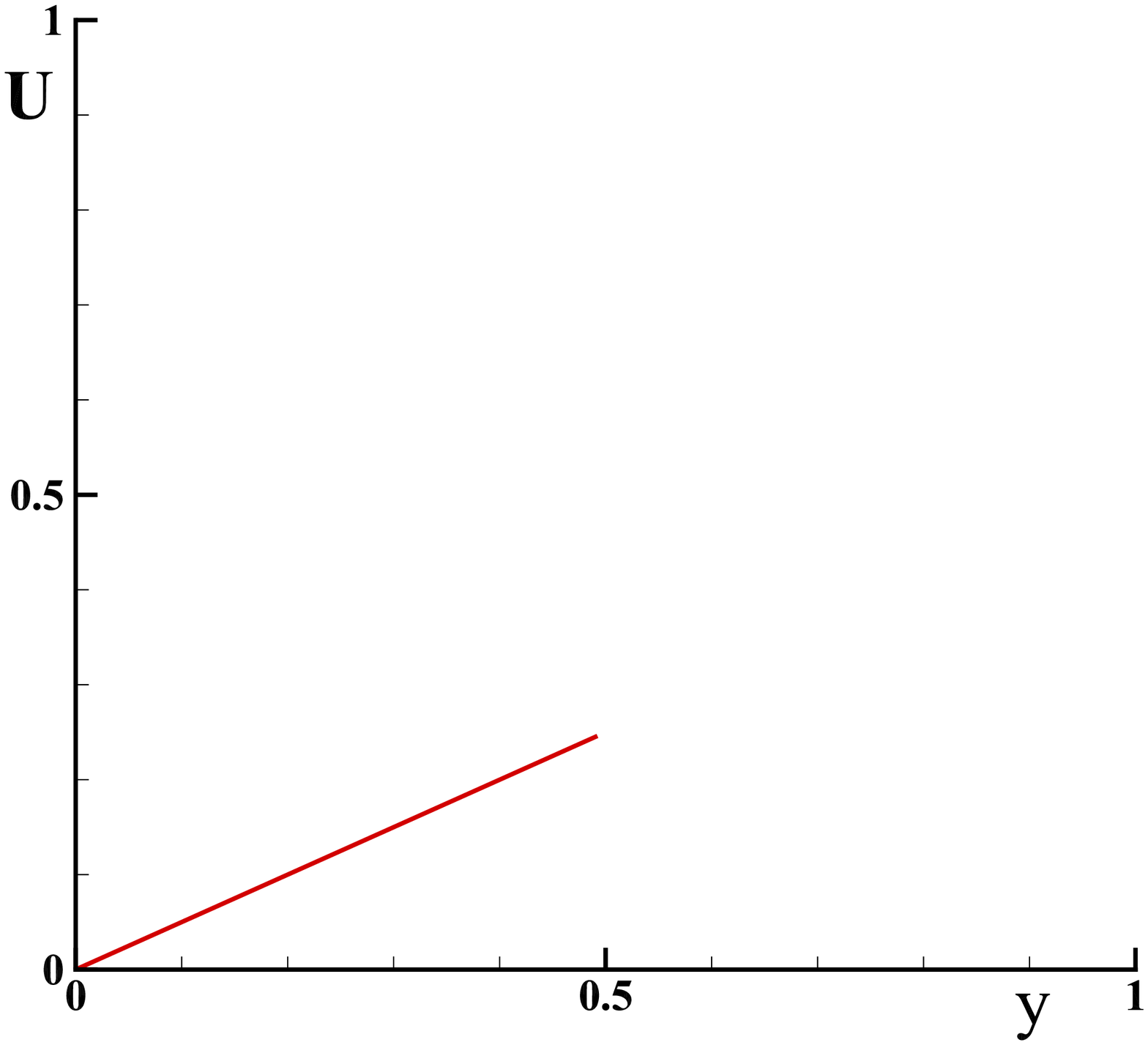}
\caption{\it Fluid-structure interaction between a viscous fluid and an elastic solid; the viscosity of the fluid is equal to $\nu = 1$ and the solid shear modulus is equal to $\nu = 4$. On the left, the velocity of the interface over time is presented, in the center, the velocity $\mathbf V$ at steady-state regime is reported and on the right, the displacement of the solid $\mathbf U$ plotted.}
\label{elas}
\end{center}
\end{figure}

Let us first consider the purely viscous case of two kinematic viscosity fluids $\nu_1 = 1$ and $\nu_2 = 4$; the solution obtained using the system (\ref{mecadis}) converges very quickly towards the stationary solution. It appears as two right-hand portions satisfying the boundary conditions and the condition $\nu_1 \: \nabla \times \mathbf V_1 = \nu_2 \: \nabla \times \mathbf V_2$ at the interface since the density on $\Sigma$ is unique. Under these conditions, the velocity at the interface is equal to $V_i = 0.2$. The 1D solution does independent on the chosen spatial approximation and the error is zero to almost machine accuracy. Note that the condition at the interface is implicitly provided by the $\nabla \times \left (\nu \: \nabla \times \mathbf V \right)$ operator. The constant rotational in each medium is respectively equal to $\nabla \times \mathbf V_1 = -1.6$ and $\nabla \times \mathbf V_2 = -0.4$. Since the problem has no compressibility terms, only the viscous terms independent of the first ones appear in the discrete motion equation.

The lower part of the domain is now assumed to behave as an elastic solid of celerity $c_t^2 = \nu = 4$. The upper part is occupied by a viscosity fluid with $\nu = 1$. The potential vector $\bm \psi^o$ makes it possible to accumulate the shear stresses in the solid, the constraints at the interface in the fluid being effectively transmitted and stored in the solid. The solution converges rapidly to a strictly zero velocity field in the solid and a linear velocity profile satisfying the condition in $y = h$ and at zero velocity on the $\Sigma$ interface. The vector equation of the system (\ref{mecadis}) is identically satisfied with $\bm \psi^o = \nu \: \nabla \times \mathbf V$ where $\mathbf V$ is the velocity of the fluid and $\bm \psi^o =  2$. The exact solution does not depend on the spatial approximation.

Figure (\ref{elas}) shows the evolution of the velocity at the $\Sigma$ interface over time. It diminishes quickly, enough to become zero. The velocity field is zero in the solid and linear in the steady-state fluid. The figure also gives the displacement $\mathbf U$ of the solid at the end of the time evolution.

While a fluid moves indefinitely under the action of shear, an elastic solid quickly reaches a stationary displacement. The absence of interpolation at the interface between a fluid and a solid allows to reach the exact solution. This very simple example makes it possible to understand the different mechanisms involved in the equation (\ref{mecadis}) and to validate the unsteady and stationary fluid-solid interaction.

In continuum mechanics, the theoretical solution of this problem can be obtained by considering the two media separately by imposing boundary conditions at the interface.
The respective equations, in stationary incompressible regime without inertial effects, are respectively for the fluid and solid media:
\begin{eqnarray}
\left\{
\begin{array}{llllll}
\displaystyle{  \nabla \cdot \left( \mu_f \: \left( \nabla \mathbf V + \nabla^t \mathbf V \right) \right) = 0 } \\  \\
\displaystyle{ \nabla \times  \left( \mu_s \: \nabla \times \mathbf U \right) = 0  }
\end{array}
\right.
\label{dissol}
\end{eqnarray}

When the properties $\mu_f$ and $\mu_s$ are constant, these equations are reduced to laplacian terms.
With the adopted assumptions, results in the fluid are obtained with the Navier-Stokes equation while solid solutions come from the Navier-Lam{\'e} equation.
The conditions at the interface are simple, for the fluid the velocity is zero at $y = h / 2$ while its value is $V_0$ at $y = h$. For the solid, the displacement is null at $y = 0$ and the constraint is imposed at the interface $y = h$, chosen equal to that of the fluid side. The velocity is of course zero in the solid domain.
The solution is very simple: $v (y) = \mathbf V \cdot e_x = \left (2 \: y / h - 1 \right) $ and $u(y) = \mathbf U \cdot e_x = \mu_f \: / \mu_s \: (2 \: y / h)$. As expected, the velocity solution $v(y)$ does not depend on the viscosity whereas the displacement depends on the ratio $\mu_f / \mu_s$.
For this simple problem, the solutions of the discrete mechanics are of course the same as in mechanics of continuous media. Among the advantages of the monolithic discrete approach, the equation of motion is unique for all media. Its acceleration formulation makes it possible to consider velocity and displacement as simple accumulators associated with operators $\nabla \cdot \mathbf V$ and $\nabla \times \mathbf V$.

\textcolor{blue}{\subsection{Extension to other constitutive laws } }

When the media have more complex rheologies, {i.e.} viscoelastic fluids, non-linear viscosity laws, viscoplastic fluids or time-dependent properties, it is possible to represent {\it a priori} their behaviour in complex situations.
In particular, the accumulation of shear-rotation constraints can only be partial and a weighting of the accumulation term of $\bm \psi^o$ by an accumulation factor $0 \leq \alpha_t \leq$ 1 makes it possible to account for viscoelastic behaviour. Threshold fluids are also easily represented by specifying the value of $\bm \psi^o = \bm \psi_c $ below which the medium behaves like an elastic solid. The rheology case with non-linear viscosities is no more a difficulty.
In fact, the discrete mechanics leads us to consider the notion of viscosity and that of rotation as attached only to the faces of the primal topology where the constraint is expressed in the form $ \nu \: \nabla \times \mathbf V $ in fluids and $ dt \: \nu \: \nabla \times \mathbf V $ in solids.

As an example, the interaction between an incompressible Newtonian viscous fluid and a Neo-Hookean elastic solid is now studied. The stress tensor expression of an incompressible isotropic hyperelastic material for the Neo-Hookean model is written as $\bm \sigma_s = - p \: \mathbf I + \mu_s \: \mathbf B $
where $\mathbf B = \mathbf F \: \mathbf F^t $ is the Cauchy-Green deformation tensor at left.
In two space dimensions, Cayley-Hamilton's theorem shows that the model of Mooney-Rivlin hyperelastic material is equivalent to the Neo-Hookean model.

The case published by K. Sugiyama in 2011 \cite{Sug11} is considered. This is a problem relating to an elastic band solicited in shear by an incompressible Newtonian fluid flow periodic in time. The laminar flow is periodic along $ x $. In the absence of the inertia terms, the problem can be solved in one dimension of space along direction $y$, with $ y \in [0, 1] $. In the present configuration, the upper interface is animated by a periodic motion $ V(t) = V_0 \: \sin (\omega \: t) $ with $V_0 = 1 $ and $ \omega = \pi $ and the lower surface is maintained at zero velocity.
\begin{figure}[!ht]
\begin{center}
\includegraphics[width=4.3cm,height=4.cm]{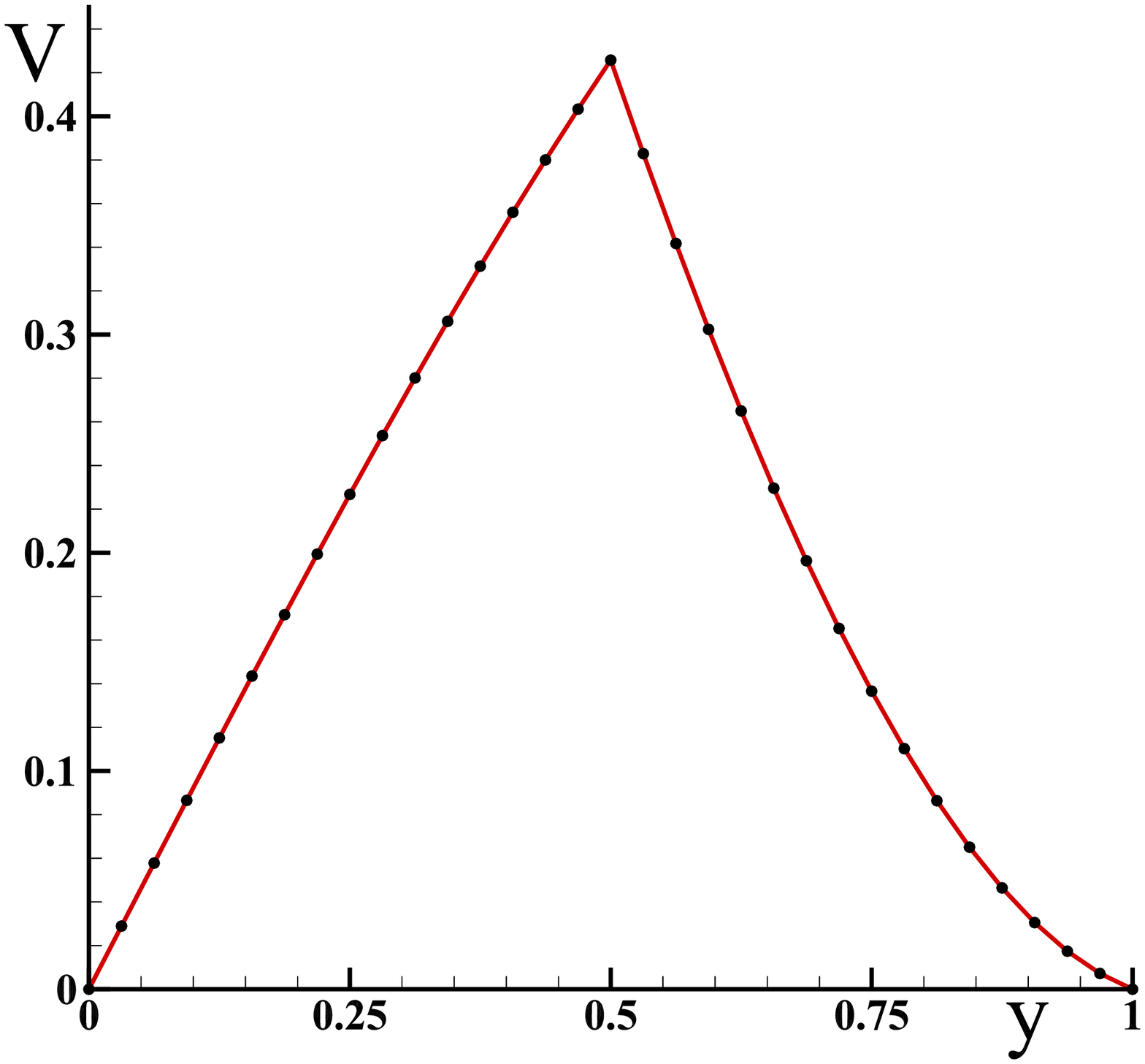}
\includegraphics[width=4.3cm,height=4.cm]{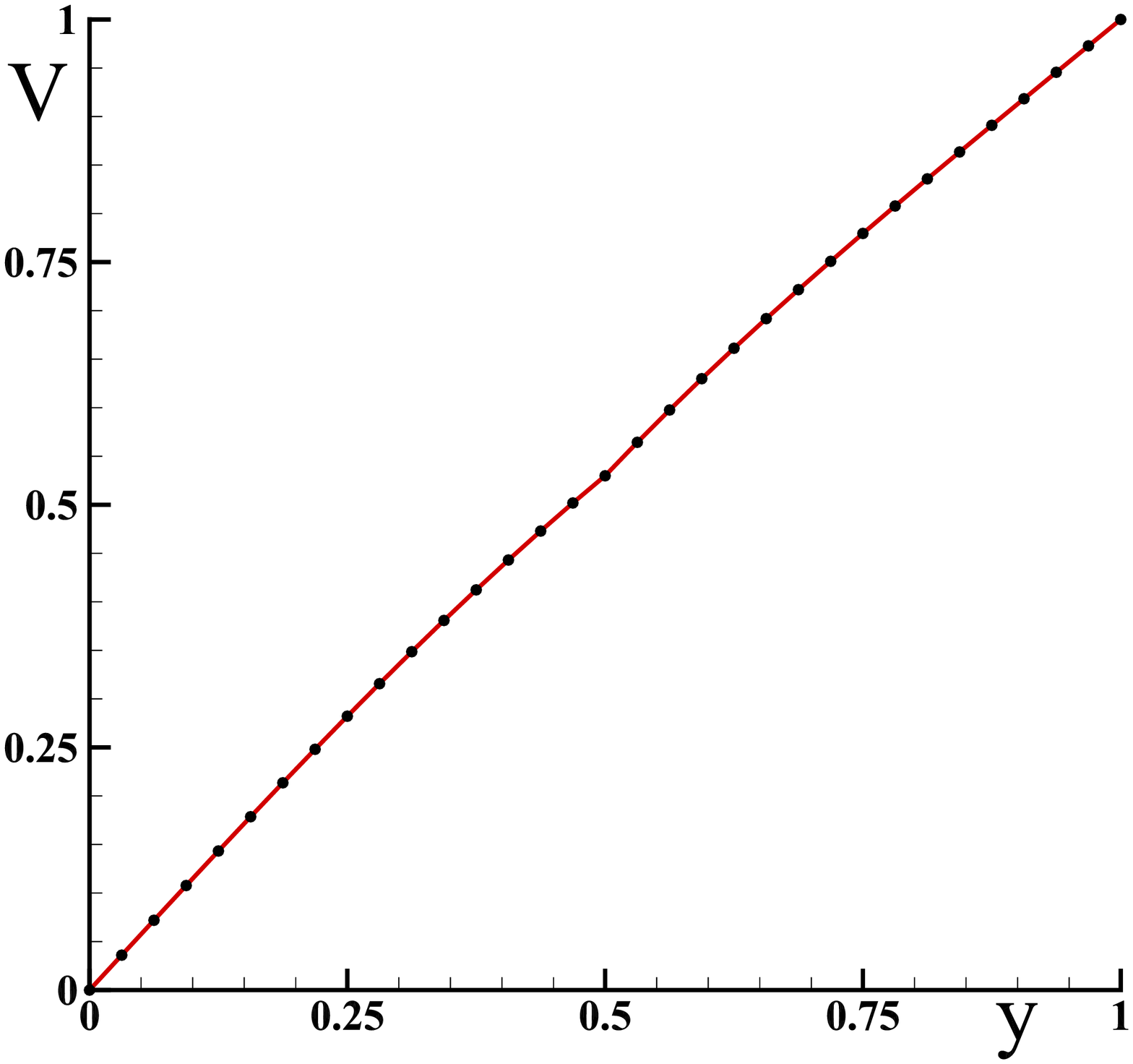}
\includegraphics[width=4.3cm,height=4.cm]{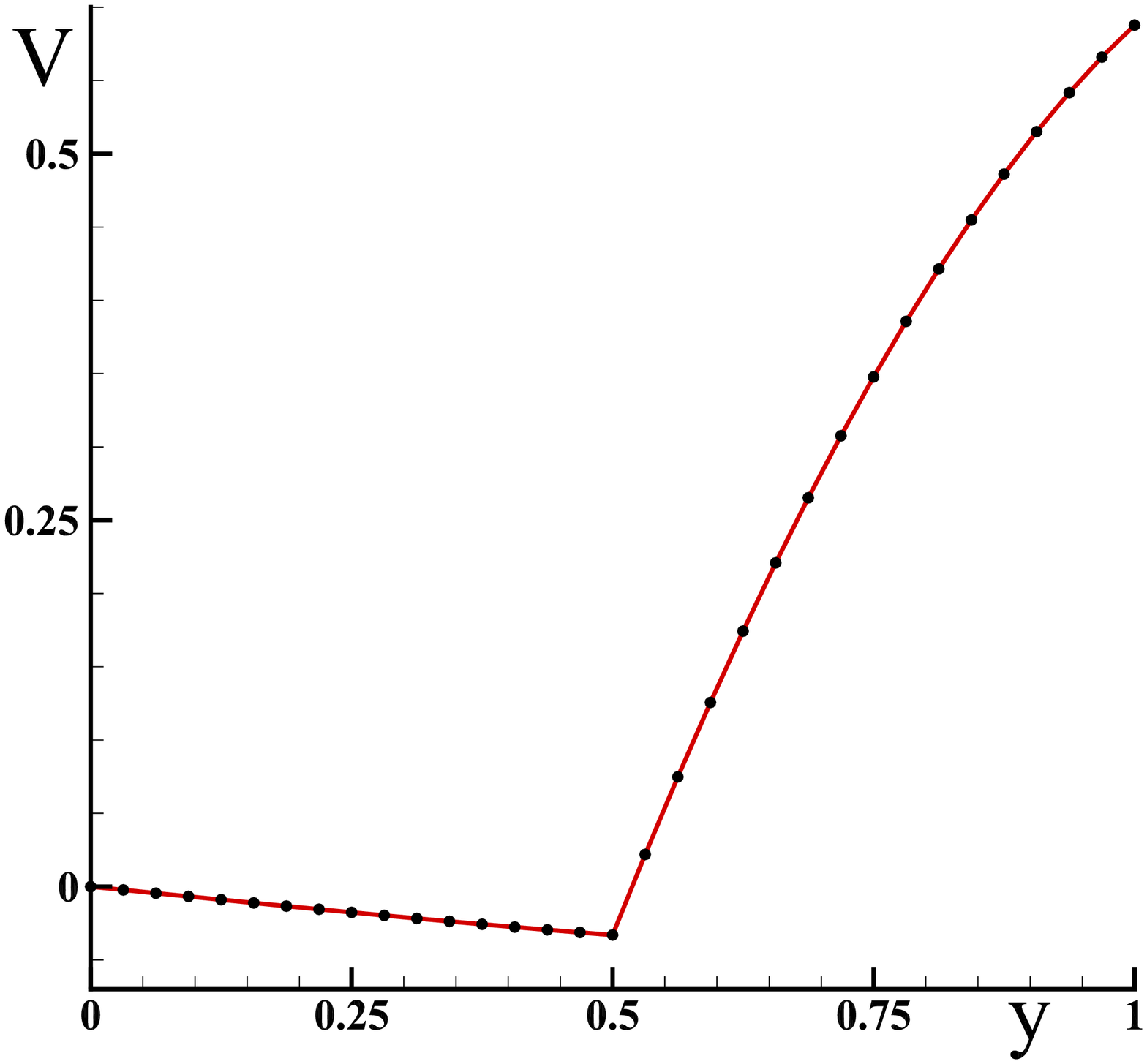}
\caption [Study of the fluid-structure interaction between a viscous fluid and an elastic solid] {\it Study of the fluid-structure interaction between a viscous fluid and an elastic solid with a periodic evolution. The viscosity of the fluid is equal to $\nu = 1$ and the solid shear modulus is equal to $\nu = 4$. Velocity profile based on $y$ for times $t = 10 \: s$, $t = 10.5 \: s$, $t = 10.8 \: s$ are plotted. In solid lines the theoretical solution is reported. The dots correspond to a spatial approximation of a $32$ mesh for $y \in [0,1]$.}
\label{sugiyamavts}
\end{center}
\end{figure}

The solid occupies the lower part of the domain while the fluid lies in the upper part, the interface between them being located at $ y =1/2$.
The theoretical solution obtained by K. Sugiyama is based on the method of variable separation applied to $y$ and time $t$.
A homogeneous solution is sought by a development based on Fourier functions in the interval $ y \in [0,1] $ and time exponential functions separately for each of the fluid and solid domains. Interfacial relations makes it possible to determine the set of the Fourier coefficients by expressing the continuity of the velocities and the stresses at the interface.
The solution $ V(y, t)$ is obtained directly by the equations of the discrete mechanics (\ref{mecadis}) for which only the conditions in $y = 0$ and $y = 1$ are imposed. Interfacial connection conditions enforcing continuity of velocity and constraint are implicitly verified by the dual rotational operator. The notion of 2D or 3D space does not exist in discrete mechanics, the operators orient the normal and tangential directions in a three-dimensional space. Despite this, in the present case, depending on the adopted assumptions, the resolution is performed in one space dimension. The chosen time step is equal to $\delta t = 10^{- 4} \: s$ in order to guarantee good overall accuracy. From the knowledge of the theoretical solution, the numerical solution is shown to behave with order two in space and time.
 
The solution is established very quickly, some periods are necessary to obtain a periodic evolution of the velocity and the profiles of the velocity are recorded starting from a time $t = 10 \: s$. The displacement of the solid is computed over time by the relation $ \mathbf U = \mathbf U^o + \mathbf V \: dt$ where $dt$ is both the differential element and the increment in time $\delta t = dt$.

Some profiles of the velocities following $y$ are given in figure (\ref{sugiyamavts}) as soon as the periodic regime is well established. A spatial and temporal convergence of order $2$ is observed. Given the absolute precision obtained (of the order of $ 10^{-4}$) with a coarse mesh $(n = 32)$, the error is unobservable on the comparison between theoretical solution and numerical simulation.

The case of fluid-structure interaction proposed by K. Sugiyama for a Neo-Hookean model has the advantage of having a theoretical solution allowing a precise validation of the numerical solutions. It also provides a basis for the development of new concepts like Discrete Mechanics. In his publication, Sugiyama gets an error in norm $L_2$ and norm $L_{\infty}$ converging with order $1$ in space whereas the DM model (\ref{mecadis}) makes it possible to reach order two with a much lower absolute error.
This very nice result is due to the separation of the properties at the interface and the absence of any interpolation in spite of an entirely monolithic and implicit treatment of the fluid-solid coupling.

Other more complex behaviour laws can be taken into account. Despite the intrinsic interest of specific studies in this area, they would not bring additional elements for the validation of the discrete model. The total disconnection between the motion equations and behaviour or state laws allows {\it a priori} to consider, as for multiphase flows for example, the use of constitutive laws of any kind.

\textcolor{blue}{\section{ Application to electromagnetism} }

Maxwell's equations are partial differential equations in time and space; it is possible to reformulate them in the exterior algebra formalism if we know how to determine the analogues of the derivation operations on the differential forms. They are also found as tensors, potentials $\phi_m$ and $\mathbf A$ or even in a form used in Special Relativity, quadrivectors.
The equation of discrete motion (\ref{mecadis}) is significantly different from these classical forms of electromagnetism. The key is to know whether this equation does not call into question the results obtained previously, but we expect significant differences in shape.

\textcolor{blue}{\subsection{ Differences and convergences} }

A fundamental difference is related to the treatment of instationarity within equations. For Maxwell's equations they are associated with the temporal variations of the electrical potential $\partial \mathbf E / \partial t$ in the Maxwell-Ampere equation and the induced magnetic field $\partial \mathbf B / \partial t$ in the Maxwell-Faraday equation; they appear as {\it ad-hoc} quantities which make it possible to represent the evolutions of these quantities but also of all those with which they are associated. In discrete mechanics these are variations of the current $d \bm j / dt$ which generate all the evolutions in time of the variables; from the physical point of view it is the variations of the acceleration which generate the strong coupling between $\mathbf E$ and $\mathbf B$. These fields become solenoidal for $\mathbf B$, $\nabla \cdot \mathbf B = 0$ and irrotational for $\mathbf E$, $\nabla \times \mathbf E = 0$ and then $\mathbf E = - \nabla e$ and $\mathbf B = \nu_m \: \nabla \times ( \bm j / \rho_m )$ where $\nu_m$ is constant on the face of the primal topology; the magnetic field induced $\mathbf B$ does not accumulate magnetic charges.

The electric field $\mathbf E$ is a polar vector defined by the gradient of the electric potential $\mathbf E = \nabla e$ while the magnetic induction field $\mathbf B$ is a pseudo-vector associated with the normal $\mathbf n$ of the primal topology; these two fields do not express themselves with the same units. In discrete mechanics or discrete electromagnetism the vectors $\nabla \phi$ and $\nabla \times \bm \psi$ are two real vectors carried by $\Gamma$ and whose sum is equal to the acceleration. The three vectors are expressed with the same units.

When there is a time dependence of electric and magnetic fields, the gauge conditions associated with Maxwell's equations are quite complex to define and apply. In discrete mechanics the two fields $\bm \gamma_{\phi} = \nabla \phi$ and $\bm \gamma_{\psi} = \nabla \times \bm \psi$ are orthogonal and therefore do not exchange anything directly, these two accelerations $\bm \gamma_{\phi}$ and $\bm \gamma_{\psi}$ are independent. The exchange mechanism is in fact complex: when one of the two fields is no longer in equilibrium with the other, it is the acceleration $\bm \gamma$ which varies and which de facto redistributes the electric currents into a magnetic field or vice versa. If $\bm \gamma = 0$ two orthogonal fields can only be locally equal to a constant, their sum is zero on a segment $\Gamma$.

Another remarkable difference is the absence of terms of inertia within Maxwell's equations; in discrete they are written $- \nabla \times (| \mathbf V |^2/2 \: \mathbf n) + \nabla (| \mathbf V |^2/2)$. It should be noted that the dual rotational operator does not exist in a continuous medium, it is similar to the gradient operator but in a direction orthogonal to it; in a discrete medium it corresponds to the circulation of an axial vector along the dual $\delta$ contour of the figure (\ref{discreet}(b)). Inertia exists whether the medium has a mass or not, as with photons for light. The variations in velocity due to inertia, those of electrons, photons and matter can be of very different values according to the cases. It should be noted that the Navier-Lamé equations do not have terms of inertia. They are practically negligible in the usual cases but that does not mean that they do not exist; the displacement of the medium must always be accompanied by inertial effects if there are spatio-temporal variations in velocity.

By combining the Maxwell-Gauss and Maxwell-Ampère equations we arrive at the law of conservation of the charge density $\partial \rho_m / \partial t + \nabla \cdot \bm j = 0$; this equation is found identically in fluid and solid. The treatment of instationarity is certainly different, but the equation of discrete mechanics has the general characteristics of Maxwell's original equations. The constraint of unifying the equations of physics requires the formulation of an equation based on a single variable, the velocity of the fluid, of the solid or the electric current.

In the absence of variations in the electric and magnetic fields, Maxwell's equations are decoupled and the equations of magnetostatic give rise to different treatments: the Coulomb approach favoring the $\phi_m$ scalar potential and the Amperian approach which favors the potential vector $\mathbf A$. In discrete mechanics, the two terms being orthogonal they must also be equal to zero separately. Similarly, in electrostatic mechanics one can define an electric potential.

\textcolor{blue}{\subsection{  Equations in terms of potentials } }

Maxwell's equations in potentials $\phi_m$ and $\mathbf A$ are obtained by combining the equations on electric potential $\mathbf E$ and magnetic induction $\mathbf B$ and, after simplification, lead to the classical relations:
\begin{eqnarray}
\left\{
\begin{array}{llllll}
\displaystyle{\nabla^2 \phi_m - \frac{1}{c^2} \: \frac{\partial^2 \phi_m}{\partial t^2} = -\frac{\rho_m}{\varepsilon_0}  } \\ \\
\displaystyle{\nabla^2 \mathbf A - \frac{1}{c^2} \: \frac{\partial^2 \mathbf A}{\partial t^2} = - \frac{\bm j}{\varepsilon_0 \: c^2}  } 
\end{array}
\right.
\label{lapphi3}
\end{eqnarray}

The second members of these equations are null in the vacuum, their solutions correspond to the propagation of a wave at $c$, the celerity of light.

Consider the case of discrete mechanics and take the divergence of equation (\ref{mecadis}). The rotational term disappears and it becomes:
\begin{eqnarray}
\displaystyle{ \nabla \cdot \bm \gamma = \nabla \cdot \left( \frac{d \mathbf V}{d t} \right) = \frac{d}{d t} \left( \nabla \cdot \mathbf V \right) + \left( \nabla \cdot \mathbf V \right)^2  } 
\label{divgam}
\end{eqnarray}

Applying this result from \cite{Cal15} to the vector equation of the system (\ref{mecadis}) assuming $c_l$ constant, leads to an equation on the scalar potential:
\begin{eqnarray}
\displaystyle{\nabla^2 \phi = -\frac{d}{d t} \left( \nabla \cdot \mathbf V \right) - \left( \nabla \cdot \mathbf V \right)^2  } 
\label{lapphi}
\end{eqnarray}
but the upgrade of $\phi$ of the same equation system makes it possible to extract $c_l^2 \: = - d \phi / dt$ and substituting:
\begin{eqnarray}
\displaystyle{\nabla^2 \phi - \frac{1}{c_l^2} \: \frac{d^2 \phi}{d t^2} = \left( \frac{1}{c_l^2} \: \frac{d \phi}{d t} \right)^2  } 
\label{lapphi2}
\end{eqnarray}

Now let us apply the primal rotational operator to the $\bm\gamma$ acceleration: 
\begin{eqnarray}
\displaystyle{ \nabla \times \bm \gamma = \nabla \times \left( \frac{d \mathbf V}{d t} \right) = \frac{d}{d t} \left( \nabla \times \mathbf V \right) + \left( \nabla \times \mathbf V \right)^2  \mathbf n  } 
\label{rotgam}
\end{eqnarray}

Considering the identity $c_t^2 \: \nabla \times \mathbf V = - d \bm \psi / dt$ and taking into account equality $\nabla \times \nabla \times \bm \psi = \nabla (\nabla \cdot \mathbf \bm \psi) - \nabla^2 \bm \psi $, as the pseudo-vector $\bm \psi$ is solenoidal $\nabla \cdot \bm \psi = 0$ , we can simplify the equation obtained and summarize the potential formulation:
\begin{eqnarray}
\left\{
\begin{array}{llllll}
\displaystyle{\nabla^2 \phi - \frac{1}{c_l^2} \: \frac{d^2 \phi}{d t^2} = \left( \frac{1}{c_l^2} \: \frac{d \phi}{d t} \right)^2  } \\ \\
\displaystyle{\nabla^2 \bm \psi - \frac{1}{c_t^2} \: \frac{d^2 \bm \psi}{d t^2} = \left( \frac{1}{c_l^2} \: \frac{d \bm \psi}{d t} \right)^2  \mathbf n } 
\end{array}
\right.
\label{lappsi2}
\end{eqnarray}

If we compare the Maxwell equations (\ref{lapphi3}) in terms of potentials $\phi_m$ and $\mathbf A$ with those of the system (\ref{lappsi2}) on $\phi$ and $\bm \psi$, a similarity is observed, but the second derivatives in time become second material derivatives. The difference is the advection of quantities at velocity $\mathbf V$; these inertial effects not taken into account in Maxwell's equations may be second order, but it cannot be denied that any medium or elementary particle is subjected to an acceleration due to variations in its velocity. Indeed, all media have finite compressibility characterized by celerities $c_l$ or $c_t$; these potential equations translate the compression energy exchanges from one point to another of a conductor. The physical significance of this phenomenon is best demonstrated in the equation of motion (\ref{mecadis}) where the equilibrium potentials $\phi^o$ and $\bm \psi^o$ represent the accumulators of the unsteady exchanges with their respective deviators $dt \: c_l^2 \: \nabla \cdot \mathbf V$ and $dt \: c_t^2 \: \nabla \times \mathbf V$.
In the first order, in the vacuum, if we neglect the inertial effects we find the expressions of the potentials directly derived from Maxwell's equations.
 Another difference is that in discrete mechanics $\phi$ and $\bm \psi$ express themselves with the same fundamental units, length and time, and above all are the potentials of the same quantity, acceleration.

In fact, the application of operators to an equation, whatever they are, inevitably leads to a loss of information; this degradation is due to the elimination of certain terms of the initial equation. The intrinsic properties of the discrete vision satisfy the two essential equalities $\nabla \times \nabla \phi = 0$ and $\nabla \cdot (\nabla \times \bm \psi) = 0 $ but it is important to keep both contributions in the same equation of motion. The resolution of this equation leads, in a single step, to the two potentials, the conservative quantity $\rho$, acceleration and velocity. The equilibrium potentials $\phi^o$ and $\bm \psi^o$ serve to represent long-term persistent quantities, magnetization, electric charge, shear stress, mechanical pressure, and so on.

\textcolor{blue}{\subsection{An example: magnetic field created by infinite length wire} }

This very simple case corresponds to a stationary phenomenon resulting from magnetostatics: a current $I$ runs through an electrical conductor of infinite length and very weak radius; it has an electrical conductivity $\sigma$, a density $\rho$ and the permeability of the external medium is equal to that of the vacuum $\mu_0$. The degeneracy of the equation of motion (\ref{mecadis}) makes it possible to obtain the equation of magnetostatics in terms of potentials:
\begin{eqnarray}
\displaystyle{ - \nabla \phi + \nabla \times \bm \psi = 0  } 
\label{fil}
\end{eqnarray}

The two quantities $\phi(x)$ and $\bm \psi(r)$ are functions of different variables and the two fields of equation (\ref{fil}) are orthogonal. The Stokes theorem and the fundamental theorem of the integral mean value make it possible to write:
\begin{eqnarray}
\displaystyle{ \int_0^{2 \pi}  \bm \psi \cdot \mathbf t \: dl = \int_a^b \nabla \phi \: dx  } 
\label{fil2}
\end{eqnarray}

To within a constant, null in this case since the lines of the magnetic field are closed, the solution of this problem is thus:
\begin{eqnarray}
\displaystyle{  \bm \psi \cdot \mathbf n  = \frac{ \left( \phi_b - \phi_a \right) }{2 \: \pi \: r} } 
\label{fil3}
\end{eqnarray}

By replacing the potentials with the usual variables of electromagnetism, $\phi = \rho_m \: e / \rho$ and $B = \rho \: \sigma \: \mu / \rho_m$ and noting that $ ( e_b - e_a) = I / \sigma$ we find the result obtained classically by the law of Biot and Savart in the form of the following component $\mathbf n$ of the magnetic field:
\begin{eqnarray}
\displaystyle{  B(r)   = \frac{ \mu_0 \: I }{2 \: \pi \: r} } 
\label{fil4}
\end{eqnarray}

The considerable interest of equation (\ref{fil}) and its solution (\ref{fil3}) bearing on the two potentials $\phi$ and $\bm \psi$, is that it is expressed only with two fundamental units, time and space. In electromagnetism the equation and its solutions involve the other fundamental units, mass M and intensity A. In the perspective of the unification of the laws of physics this would remain a difficulty; indeed, it is not an analogy that is sought; it is a unique equation. 

\textcolor{blue}{\subsection{Magnetic field in a torus} }

Many cases of practical interest are inspired by the design of electric motors and, in general, machines where magnetic fields and electric fields interact. The case treated here does not refer to an industrial problem, it highlights the properties of the discrete equation on a simple problem: a conductive coil is considered, made of a copper toroid, traversed by an electric current $\mathbf I$ inducing a magnetic field $\mathbf B$ in the surrounding medium. The near field can be obtained by integrating the equations of the magnetostatic physics $B = \mu \: I \: R^2 / (2 (R^2 + z^2)^(3/2))$ where $R$ is the radius of the turn and $z$ is the coordinate orthogonal to the plane of the torus.

\begin{figure}[!ht]
\begin{center}
\includegraphics[width=6.cm,height=4.cm]{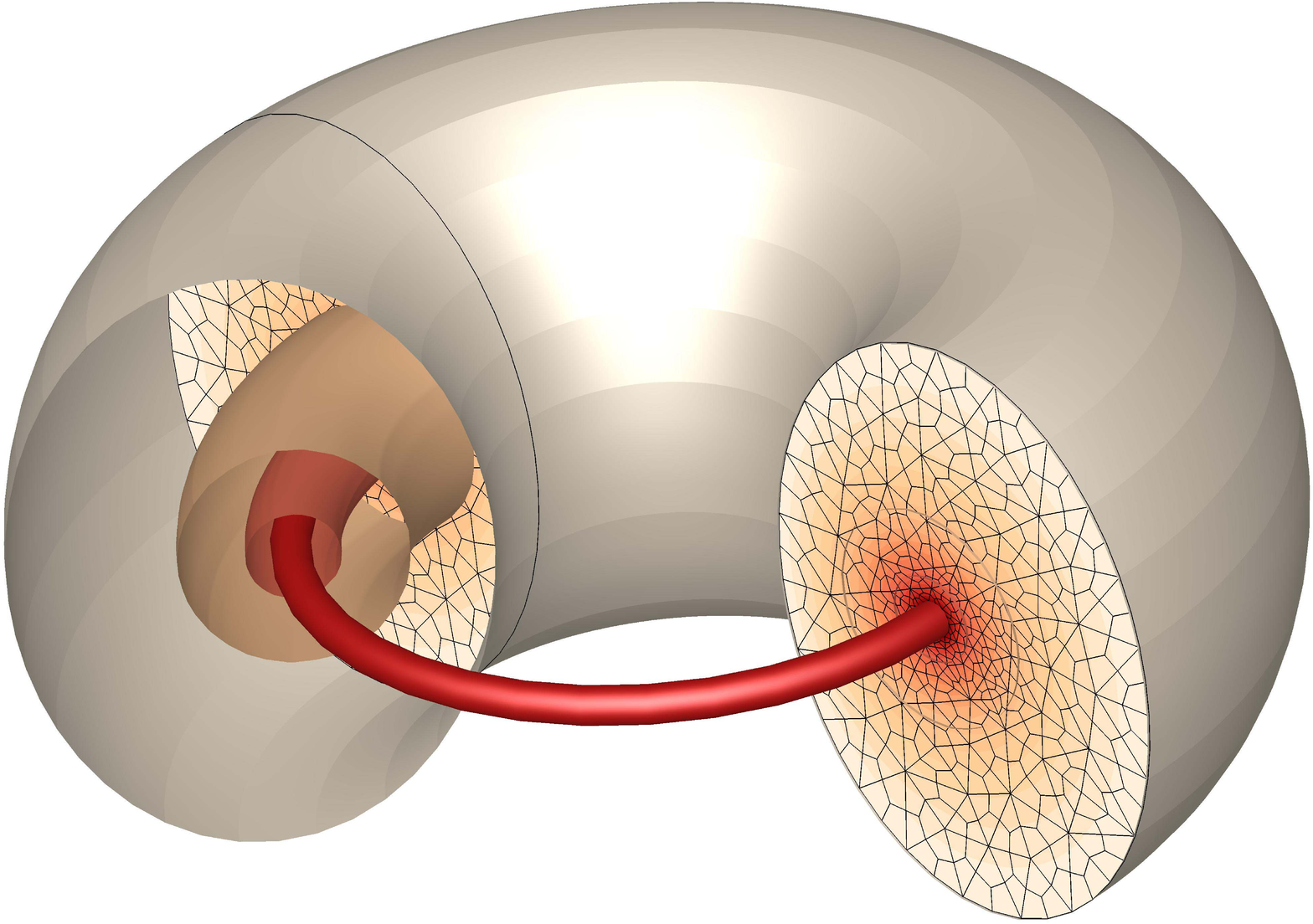} 
\includegraphics[width=3.cm,height=4.cm]{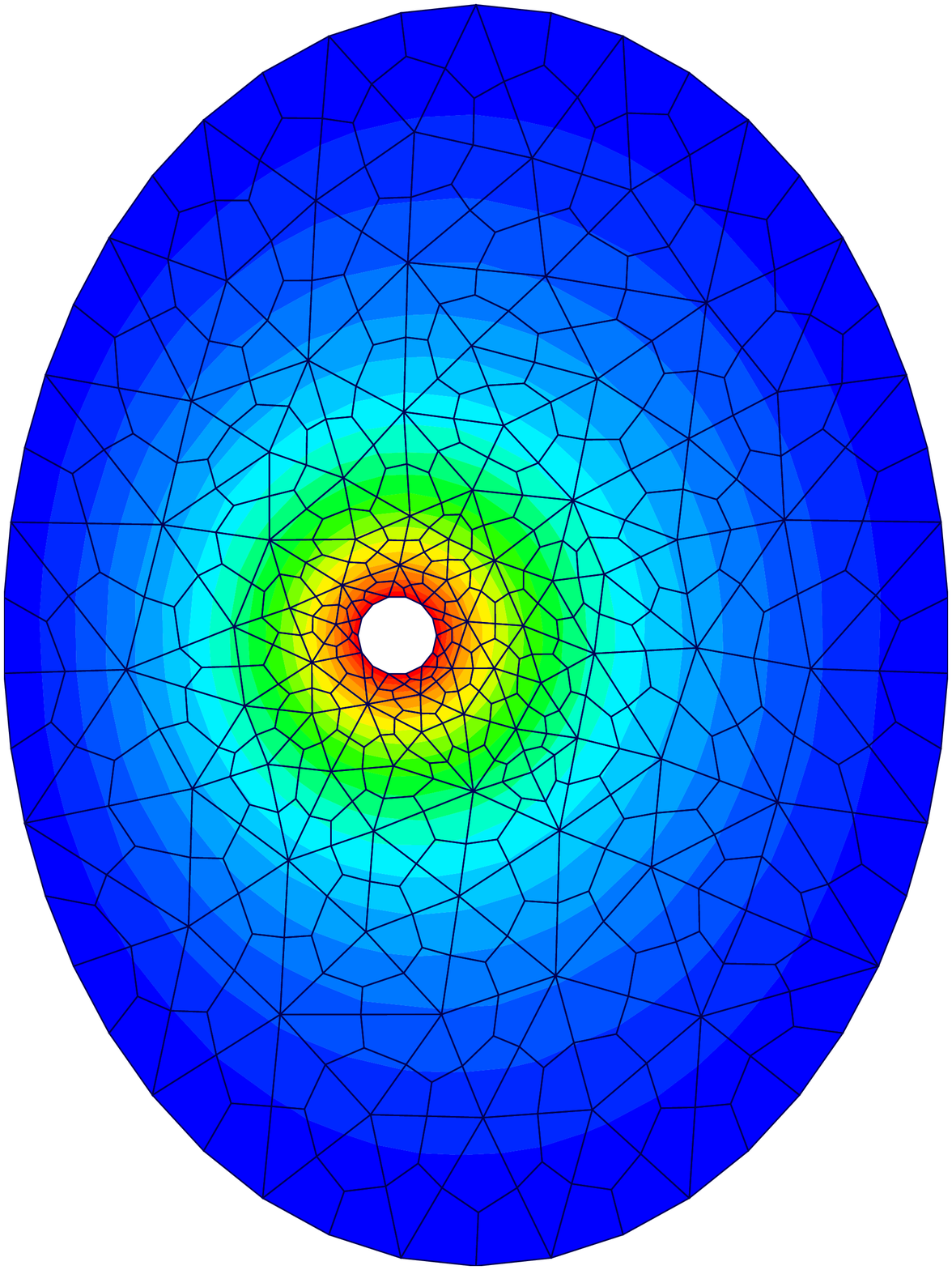} 
\vspace{-1.mm}
\caption{\it Magnetic field created by a circular turn crossed by a current $I$ within a torus of elliptical section. The unstructured mesh is composed of non-regular hexahedra \cite{Guez09}. On the left is the torus coil surrounded by isovalues of the magnetic field and on the right, a cross-section representing the potential field is shown. }
\label{tomawak}
\end{center}
\end{figure}

The problem is simulated by assuming that the turn is contained in a torus of elliptical section delimiting a zero electric field surface. The three-dimensional domain is meshed with {\it gmsh} \cite{Guez09} in the form of an unstructured tessellation with a reduced number of cells conforming to the toric surfaces. Figure (\ref{tomawak}) illustrates the geometry used and the unstructured mesh adopted.

  From the values of the electrical potential $e$ at the ends of the turn, the properties of the media and the geometry of the chosen case, it is possible to define the scalar potential $\phi$ and to deduce the current density $\bm j$ to be imposed, represented by the velocity $\mathbf V$ in the discrete model. As the current is imposed, the resulting gradient of scalar potential is a constant $\nabla \phi^o = Cte$ and the equation to be solved becomes:
\begin{eqnarray}
\left\{
\begin{array}{llllll}
\displaystyle{ \frac{\partial \mathbf V}{\partial t} + \nabla  \left( \frac{| \mathbf V |^2 }{2}  \right) - \nabla \times \left( \frac{| \mathbf V |^2 }{2} \: \mathbf n \right)  + \nabla \times \left( \nu_m \: \nabla \times \mathbf V \right) = - \nabla  \phi^o  } \\  \\
\displaystyle{ \bm \psi  =  - \nu_m \: \nabla \times \mathbf V }
\end{array}
\right.
\label{tomawac}
\end{eqnarray}

At the end of the unsteady process, the obtained solution relates to the only component $\mathbf V$ of the velocity on each of the segments $\Gamma$ of the primal topology. The magnetic field $\bm \psi$ is obtained at every instant by an upgrade from $\nabla \times \mathbf V$. The magnetic field lines are almost circular. They are contained in each elliptical section orthogonal to its main axis. Figure (\ref{tomawak}) shows the electric flux tubes around the coil. It should be noted that the electric field exists in the whole field, in the copper turn as well as within the elliptical core. On the contrary, the electric charge density exists only in the turn. This is actually the $\mathbf V = \bm j / \rho_m$ velocity that is searched for and this ratio always keeps a physical meaning, even in the limit when $\rho_m \rightarrow 0$ in the absence of current.

If we neglect the inertial effects, the equation reaches a steady state for which $\nabla \times (\nu_m \: \nabla \times \mathbf V) = - \nabla \phi^o$. We consider that $\nu_m$ is a constant and that the field $\mathbf V$ is solenoidal. Under these conditions, the equation (\ref{tomawac}) becomes a simple vectorial Poisson equation $\nu_m \: \nabla^2 \bm \psi = \nabla \phi^o$. For this example, the scalar field $\psi = \bm \psi \cdot \mathbf n$ depends on two space variables and so is the same whatever the plane defined by an elliptical cross section of the external torus.

This simulation of an induced magnetic field problem deals with a stationary case with obvious symmetries but it shows the versatility of the equation of discrete motion. The problem has been solved with an unsteady formulation of the vectorial equation (\ref{tomawac}) and the solution deals with the quantities $(\mathbf V, \phi, \bm \psi)$. Since the relations between these quantities and the classical variables of electromagnetism are bijective, it is of course possible to recover these variables, even if it is not a necessity. In fact, each field of physics has its variables, its physical properties, nevertheless they are not all independent. The discrete formulation presented here not only unify mechanics and electromagnetism, but also proposes unique variables for these domains. These variables, which are expressed only with two fundamental units, as well as the physical properties involved in the formulation, could possibly be extended to other areas of physics.

\vspace{2.mm} 
 In the general framework of electromagnetism it is the discrete equation (\ref{mecadis}) that must be integrated directly into space and time; this makes it possible to find the solution on the velocity $\mathbf V$ by knowing its value at the previous time $t^o$. Any disturbance of the magnetic or electric field, any variation in a source term or boundary condition generates an acceleration $\bm \gamma$ which extends to both components of its Hodge-Helmholtz decomposition. Understanding the behaviour of this discrete equation is very complex and it would be illusory to seek to explain it through trivial reasoning. It is only necessary to find the results obtained conventionally with laws of physics established previously, which does not mean that the equations must be identical. For example, the equation of discrete mechanics is significantly different from the Navier-Stokes equation and yet the solutions are the same. Although different from one of Maxwell's forms, the equation of discrete movement must serve to find the same results as in electromagnetism. The formalism presented to unify certain equations of physics cannot escape the concepts of General Relativity introduced by Einstein, encompassing those of Newton and many others before him, including Galileo.

\textcolor{blue}{\section{Conclusions} }

Each field of physics has its equations and variables: pressure, stress and velocity for fluids, compression and shear stresses and displacement  for solids or electrical potential, magnetic field and current for electromagnetism. Unification, if possible, requires the definition of common quantities and the return to specific variables is not a necessity. For example, the potential $\phi = p / \rho_f$ in fluid can be used for solving the equation and for the subsequent use of the results without having to go back to the pressure, even if it is possible.

The system equation (\ref{mecadis}) considered as generic has remarkable properties induced by its form, the two components of the Hodge-Helmholtz decomposition of acceleration. The terms potential sources, gravity and capillary effects can also be written following the same principle. The boundary conditions will be written in the same way by the introduction of irrotational and solenoidal terms into the equation. The four properties, the factors $\alpha_l$ and $\alpha_t$ and the celerities $c_l$ and $c_t$ can be constant or a function of the variables themselves, they can only be fixed by the experiments.
It is noticeable that all quantities are expressed only with two fundamental units, those of length and time.
It is remarkable that it is the generic law $\bm \gamma = \mathbf g$ resulting from Galileo's concepts which is at the origin of this.




\end{document}